\documentclass[%
 aip,
 floatfix,
 reprint,%
 author-year,%
]{revtex4-1}

\usepackage{lmodern}
\usepackage{graphicx}
\usepackage{amsmath,amssymb,amsfonts,mathrsfs,bm}
\usepackage{mathtools}
\usepackage[labelformat=simple]{subcaption}
\usepackage{siunitx}

\newcommand{\rd}{\mathrm{d}}

\newcommand{\re}{\mathrm{e}}
\newcommand{\Th}{\mathrm{Th}}
\newcommand{\Te}{\mathrm{Te}}
\DeclareMathOperator{\Beta}{Beta}
\DeclareMathOperator{\HalfNormal}{HalfNormal}
\DeclareMathOperator{\ExpDist}{Exp}
\DeclareMathOperator{\Laplace}{Laplace}

\makeatletter
\newcommand*{\balancecolsandclearpage}{%
  \close@column@grid
  \cleardoublepage
  \twocolumngrid
}
\makeatother

\usepackage{hyperref}
 \hypersetup{
     colorlinks=true,
     linkcolor=blue,
     filecolor=blue,
     citecolor=blue,      
     urlcolor=blue
}

\allowdisplaybreaks

\begin{document}

\title{Hierarchical Bayesian inference for uncertainty quantification of thermal grease rheology}

\author{Pranay P. Nagrani}
\affiliation{School of Mechanical Engineering, Purdue University, West Lafayette, Indiana 47907, USA}

\author{Akshay J. Thomas}
\affiliation{School of Mechanical Engineering, Purdue University, West Lafayette, Indiana 47907, USA}
\affiliation{Composites Manufacturing and Simulation Center, Purdue University, West Lafayette, Indiana 47906, USA}

\author{Amy M. Marconnet}
\affiliation{School of Mechanical Engineering, Purdue University, West Lafayette, Indiana 47907, USA}

\author{Ivan C. Christov}
\thanks{Author to whom correspondence should be addressed}
\email{christov@purdue.edu}
\affiliation{School of Mechanical Engineering, Purdue University, West Lafayette, Indiana 47907, USA}

\date{\today}

\begin{abstract}
Rheologically complex soft solids such as thermal greases consist of filler particles within a polymer matrix. These materials find applications in improving the conformity of solid-solid contacts and enhancing heat transfer. Complex soft solids exhibit a transient non-Newtonian rheological response, including thixotropy and viscoelasticity. Previously, stress relaxation and buildup in sheared commercial thermal greases were successfully captured using a nonlinear elasto-visco-plastic (NEVP) model and a thixo-elasto-visco-plastic (TEVP). However, the previous model calibration methods ignored parameter uncertainty, providing only single values of the rheological parameters, and did not quantitatively address the chosen model's identifiability from the data or credibility of the calibration. We address these limitations via hierarchical Bayesian inference, accounting for uncertainties arising from epistemic and aleatoric sources. Importantly, the hierarchical approach allows us to assimilate experiments measuring the stress responses at various startup shear rates by allowing the models' parameters to vary across different shear rates. Then, a global distribution and the associated uncertainty are obtained by pooling. We also propagate uncertainties to the transient shear stress response predicted by the models. Overall, we demonstrate that the chosen NEVP and TEVP models are identifiable from rheometric startup data. However, for the TEVP model, the uncertainty of the parameters is lower (narrower distributions) when higher shear rates are used for inference.
\end{abstract}


\maketitle

\section{Introduction}
\label{sec::Intro}
Thermal greases are soft solid materials composed of filler particles within a polymer matrix \citep{Prasher2002RheologicalExperimental,Prasher2002RheologicalModeling}. Thermal greases find applications at solid-solid interfaces within electronics packages, where they improve contact and, thus, enhance heat transfer. Such materials are often subjected to deformations due to warpage of solid substrates, resulting in applied stresses that can cause flow. Due to their microstructure, thermal greases display rheological behaviors beyond viscous resistance, such as elasticity and thixotropy \citep{Nagrani2023Data-drivenGreases,Prasher2002RheologicalModeling}. Understanding the rheology of thermal greases can help design better thermal interface materials with improved performance (i.e., lower thermal resistances) and reliability \citep{Prasher2006ThermalDirections,Nagrani2023Data-drivenGreases,Due2013ReliabilityReview}. 

Previous research described the thermal grease rheology using steady-state viscoplastic models, such as the Bingham and Herschel--Bulkley (HB) equations, and correlated the greases' thermal resistance to the rheological parameters of these models \citep{Prasher2001SurfaceMaterials,Prasher2002RheologicalExperimental,Prasher2003ThermalMaterials,Prasher2006ThermalDirections}. They found that the thermal resistance of these greases is directly proportional to the bond line thickness (BLT), which is the final thickness obtained after squeezing the soft material between two solid surfaces at constant pressure \citep{Ramakrishnan2021CPUCooling,Prasher2006ThermalDirections}. \citet{Prasher2003ThermalMaterials} showed that the BLT is a strong function of the thermal grease rheology and the model chosen to represent it. Specifically, Newtonian and shear-thinning power-law rheological models inaccurately predict the final BLT to be $0~\si{\micro\meter}$ upon squeezing, while the Bingham and HB models capture the finite nonzero thickness. Rheological characterization also helps to explain the degradation of thermal greases subjected to thermal cycling. For example, \citet{Nagrani2023InfluenceGreases} proposed the dynamic apparent viscosity as a function of temperature as a metric to predict the microstructural changes in a thermal grease and their effect on its performance against thermal stresses induced by thermal cycling at fixed BLTs. 

These prior works only consider the steady rheological response of a thermal grease and fit the stress as a function of the rate of shear strain, reporting a single best-fit value of each rheological parameter. However, the complex microstructure rearrangement of these amorphous soft solids leads to uncertainties in their rheological characterization from steady-state measurements \citep{Prasher2006ThermalDirections}. Different imposed stress conditions induce different degradation rates as a result of the rearrangement of the grease's microstructure. These degradation rates must be captured in the rheological model to understand the uncertainty in the bulk properties \citep{Due2013ReliabilityReview,Jiang2022TheoreticalEnvironment}. Therefore, the goal of the present work is to develop a probabilistic framework using hierarchical Bayesian inference to enable the calibration of more advanced viscoelastic and thixotropic rheological models of thermal greases and quantify the uncertainties of the inferred rheological parameters of these thermal greases.
 
Probabilistic tools, such as Bayesian inference \citep{Gelman2013,Albert2019}, have been widely used for model calibration and uncertainty quantification in various applications, ranging from heat transfer \citep{Wang2004AProblem,Kaipio2011TheTransfer,Parthasarathy2008EstimationPriori,Mousavi2025BayesianAssimilation} to fluid mechanics \citep{Cotter2009BayesianMechanics,Kim2019,Blanchard2021}, transport phenomena \citep{Lysy2016}, hemodynamics \citep{Paun2020}, acoustics \citep{Juniper2022}, solid mechanics \citep{Rappel2020AMechanics,Rappel2019IdentifyingUncertainty,Thomas2022BayesianComposites}, and rheology \citep{Shanbhag2010,Freund2015,Freund2018,Paul2021,Rinkens2023,Ran2023UnderstandingInference,Mangal2025}, to name a few. Specifically, in rheology, the Bayesian approach has been used to estimate the uncertainty of calibrated model parameters obtained by inverse methods, as well as for assessing the identifiably and ``credibility'' of the inferred models \citep{Freund2015}. 

For example, \citet{Rinkens2023} used a Bayesian approach to calibrate a truncated power-law rheological model for a shear-thinning polymeric solution. They highlighted how the Bayesian approach incorporates new rheometric data and recalibrates the model, as well as reveals features of the flow in the chosen squeezing configuration that the experimental apparatus cannot assess. On the other hand, \citet{Ran2023UnderstandingInference} performed the uncertainty of rheological model calibration for kaolinite clay suspensions (a type of ``mud''), which exhibited thixotropy and viscoelasticity much like the present thermal greases, focusing on uncertainty arising due to variations in volume fractions. Bayesian inference revealed that the mud is strongly viscoelastic and weakly thixotropic at low filler particle volume fractions and purely thixotropic and inelastic at high volume fractions. Meanwhile, \citet{Rappel2019IdentifyingUncertainty} quantified the influence of model error (in addition to errors in strain and stress measurements) in the calibration of rheological models for elastic and elastoplastic solids.

Due to the presence of numerous models that can explain the rheological data, \citet{Freund2015} motivated the need for Bayesian inference methodologies for model selection. From a survey of constitutive models calibrated to oscillatory simple shear rheological data for aqueous polyvinyl alcohol with sodium tetra-borate (PVA-borax) network and a gluten gel, they showed how a Bayesian approach demonstrates that models ``grounded" in physics have greater generalizability than empirical models, which may have a larger number of fitting parameters. Motivated by these prior works, here, we leverage the hierarchical Bayesian inference to quantify the uncertainties of the model parameters inferred from transient rheometric measurements of thermal greases subjected to suddenly imposed shear. 

Toward this same end, previously, \citet{Nagrani2023Data-drivenGreases} introduced an approach based on physics-informed neural networks (PINNs) \citep{Raissi2019}. PINNs enable parameter estimation by regularizing the likelihood of the data with prior information encoded via differential equations. PINNs were leveraged to calibrate model parameters of empirical rheological models describing the stress buildup and stress relaxation regimes. However, PINNs only estimate a single value for the parameters. Thus, this prior approach did not obtain the distribution densities of the calibrated model parameters.

The objective of this work is to address this limitation by using Bayesian inference to characterize the uncertainty of the rheological model parameters of thermal greases in the stress buildup and relaxation regimes. Importantly, improving on previous studies employing Bayesian inference in rheology, we take a \emph{hierarchical} approach, which allows for model parameters to vary across different shear rates for which experimental data was obtained \citep{Lysy2016}. Such a variation would be impossible to capture using standard Bayesian models; we would obtain different parameter values and distributions for different shear rates, or we would obtain one set of parameters for all shear rates, representing two extremes. We further propagate the uncertainty to transient shear stress profiles to understand the identifiability of the models and their credibility in the stress buildup and relaxation regimes. Our framework naturally accounts for the epistemic (due to limited availability of experimental data) and aleatoric (due to the random nature of rheometric experiments) uncertainties observed in the shear stress profiles, adding to the list of success of data-driven rheology \citep{Mangal2025}.

To this end, the remainder of the paper is organized as follows. Rheological experiments and two suitable models for the stress evolution are introduced in Sec.~\ref{sec::CM_expt}. Next, we propose data-generating models for the experiments conducted and discuss methods for estimating the posterior distributions of the parameters of interest in Sec.~\ref{sec::BI}. The uncertainty quantification of the model parameters and propagation to the shear stress profiles is discussed in Sec.~\ref{sec::RAndD}. Finally,  Sec.~\ref{sec::Conclusion} concludes the study.

\section{Rheological experiments and stress evolution models}
\label{sec::CM_expt}

\begin{figure*}
    \centering
    \begin{subfigure}[t]{0.47\linewidth}
      \includegraphics[width=\linewidth]{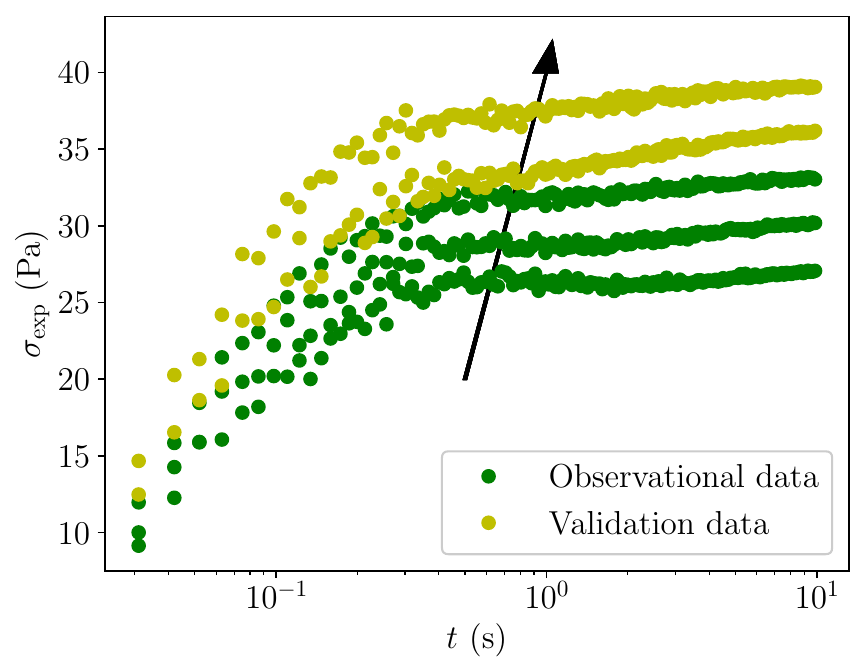}  
      \caption{}
      \label{fig:data_nevp}
    \end{subfigure}
    \qquad
    \begin{subfigure}[t]{0.47\linewidth}
      \includegraphics[width=\linewidth]{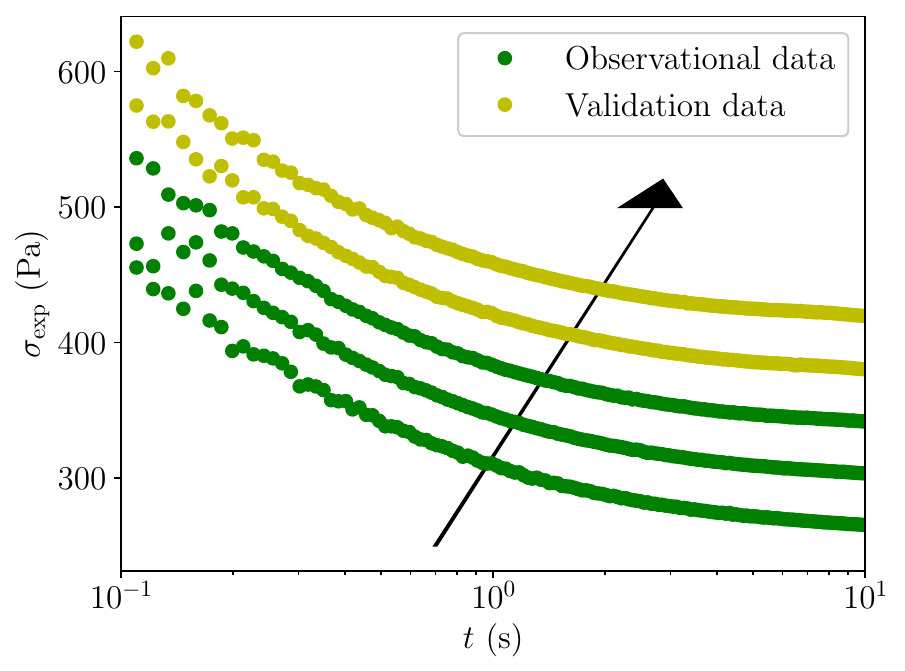}  
      \caption{}
      \label{fig:data_tevp}
    \end{subfigure}
    \caption{Experimental stress measurements, $\sigma_\mathrm{exp}$, from \citet{Nagrani2023Data-drivenGreases} based on startup rheometric experiments. (a)~Data for DOWSIL\textsuperscript{TM} TC-5550 at shear rates $\dot{\gamma}_0 = 0.06~\si{\per\second}$ (bottom curve), $0.07$, $0.08$, $0.09$, and $0.1~\si{\per\second}$ (top curve) illustrating transient stress buildup. (b)~Data for DOWSIL\textsuperscript{TM} TC-5622 at shear rates  
    $\dot{\gamma}_0 = 6~\si{\per\second}$ (bottom curve), $7$, $8$, $9$, and  $10~\si{\per\second}$ (top curve) illustrating stress relaxation. Each curve corresponds to an experiment at fixed $\dot{\gamma}_0$, and the arrows indicate the direction of increasing $\dot{\gamma}_0$. The darker points are the data used for calibrating the uncertainties in the models' parameters, while the lighter points are used to validate the models' predictions.}
    \label{fig::expt_data}
\end{figure*}

We characterize the uncertainty of the rheological parameters inferred from two different rheological models based on existing experimental data in two different regimes --- stress buildup and stress relaxation --- observed in the DOWSIL\textsuperscript{TM} TC-5550 and  DOWSIL\textsuperscript{TM} TC-5622 thermally conductive compounds, respectively. Rheological data for these commercially available thermal greases was collected in our previous study \citep{Nagrani2023Data-drivenGreases}. Technical data sheets \citep{Dow2017DOWSILSheet,Dow2022DOWSILSheet} only provide a single value of the material properties for the thermal greases and do not report any uncertainty metrics. The composition of these soft solids, such as the type of filler particle, volume fraction, or diameter, is unknown \textit{a priori}, necessitating that the rheological model infer the effects of the microstructure from only macroscopic measurements. Indeed, our initial rheological experiments showed that  DOWSIL\textsuperscript{TM} TC-5550 and DOWSIL\textsuperscript{TM} TC-5622 exhibit non-Newtonian behavior; specifically, they possess significant elasticity and yield stress, as is typical of amorphous soft solids \citep{Coussot2007RheophysicsApproaches,Bonn2017}. These considerations allowed us to narrow down the possible empirical rheological models to describe these thermal greases' rheology \citep{Nagrani2023Data-drivenGreases}. We first summarize the previous rheological experiments in which we observed stress buildup or relaxation within these materials. Then, we review the relevant rheological models used to describe the observed stress evolution. 

\subsection{Rheological data}
\label{sec:rheology_expt}

\citet{Nagrani2023Data-drivenGreases} performed rheological experiments on the transient stress evolution in thermal greases to gather data for the calibration of rheological models. In this study, we use the openly available data of \citet{Nagrani_PINNs_rheology}. The reader is referred to these previous works for the experimental details. In short, the thermal greases were subjected to a pre-shear protocol to erase history effects. Then, they were subjected to a step-strain startup protocol, such that $\dot{\gamma}(t) = H(t)\dot{\gamma}_0$, where $\dot{\gamma}_0$ is the imposed constant shear rate, and $H(t)$ is the Heaviside unit step. The shear stress $\sigma(t)$ was measured as a function of time, $t$, for different imposed shear rates. 

The evolution of the experiment shear stress $\sigma_\mathrm{exp}(t)$ for different imposed shear rates $\dot\gamma_0$ is shown in Fig.~\ref{fig:data_nevp} for DOWSIL\textsuperscript{TM} TC-5550, where the subscript ``exp'' stands for experimental measurements. This thermal grease exhibits stress buildup that levels out towards a steady state at late times. On the other hand, DOWSIL\textsuperscript{TM} TC-5622 displays stress relaxation, as seen in Fig.~\ref{fig:data_tevp}, with a decrease in stress over time. From the stress response at five different shear rates, three are chosen as ``observational data'' for calibration, and two are used as ``validation data.'' Since extrapolation is a more challenging aspect for any model (in comparison to interpolation), we use the observational data from the lower shear rate to predict the uncertainties in the shear stress response at the higher shear rates. Then, we compare the results to the validation data.

\subsection{Rheological models and their solutions}
\label{sec:cons_models}
Since the two thermal greases display vastly different rheological responses, two models are considered: the nonlinear elasto-visco-plastic (NEVP) model (Section~\ref{sec:NEVP}) for the DOWSIL\textsuperscript{TM} TC-5550 grease and a thixo-elasto-visco-plastic (TEVP) model for the DOWSIL\textsuperscript{TM} TC-5622 grease (Section~\ref{sec:TEVP}).

\subsubsection{Nonlinear elasto-visco-plastic (NEVP) model}\label{sec:NEVP}
We use the NEVP model of \cite{Kamani2021} to describe the stress buildup regime. This rheological model unifies the physics of yield stress fluids before and after yielding. At a constant shear rate (i.e., $d\dot\gamma/dt=0$), the model takes the form
\begin{equation}
      \label{eq:NEVP}
      \frac{d\sigma}{dt}  = \frac{G\dot\gamma(t)}{\sigma_y+K\dot\gamma(t)^{n}+\eta_s\dot\gamma(t)} \big\{\sigma_y + K \dot \gamma(t)^n -\sigma(t) \big\},
\end{equation}
where $G$ is an elastic shear modulus, $\eta_s$ is the solvent viscosity, and $\sigma_y$ is the yield stress. The power-law exponent $n$ and the consistency index $K$ are understood in the sense of an HB fit; that is, at steady state ($d\sigma/dt=0$),  $\sigma = \sigma(\dot{\gamma}) = \sigma_y + K \dot\gamma^{n}$. In this model, the unknown parameters whose uncertainty is to be calibrated are $G$, $\eta_s$, $\sigma_y$, $K$, and $n$.

Recently, \citet{Zhang2025Data-drivenConduction} observed that Eq.~\eqref{eq:NEVP} can be solved analytically for certain startup protocols. Specifically, for the case of sudden step loading (startup rheometric protocol) such that $\dot{\gamma}(t) = H(t)\dot{\gamma}_0$, the NEVP is actually a linear ordinary differential equation (ODE). We can take the Laplace transform $\mathcal{L}$ of Eq.~\eqref{eq:NEVP} and solve for $\mathcal{L}\{\sigma(t)\}(s) \equiv \bar{\sigma}(s) = \int_0^\infty \sigma(t) \re^{-st} \,\rd t$, to obtain:
\begin{multline}
    \bar{\sigma}(s) = \frac{1}{s + \frac{G\dot\gamma_0}{\sigma_y+K\dot\gamma_0^{n}+\eta_s\dot\gamma_0}} \\ 
    \times \left\{ \sigma(0) + \frac{G\dot\gamma_0}{\sigma_y+K\dot\gamma_0^{n}+\eta_s\dot\gamma_0} \left[\sigma_y + K \dot \gamma_0^n \right] \frac{1}{s} \right\}.
\end{multline}
The Laplace inversion is easily performed using tables of inverses, to obtain
\begin{multline}
    \sigma(t) = \sigma_y + K \dot \gamma_0^n + [ \sigma(0) - \sigma_y - K\dot{\gamma}_0^n ] \\
    \times \exp\left(-\frac{G\dot\gamma_0}{\sigma_y+K\dot\gamma_0^{n}+\eta_s\dot\gamma_0}t\right).
    \label{eq:NEVP-sol-step-strain}
\end{multline}
Observe that when $\sigma(0) < \sigma_y + K\dot{\gamma}_0^n$, the stress will grow towards $\sigma_y + K \dot{\gamma}_0^n$ as $t\to\infty$, which corresponds to stress buildup. 

\subsubsection{Thixo-elasto-visco-plastic (TEVP) model}\label{sec:TEVP}
During stress relaxation, the material flows, causing complex motion of the filler particles within the polymer matrix. These microstructural rearrangements lead not only to viscous resistance and elasticity but also \emph{thixotropy}, i.e., time-dependent material evolution. Thus, the NEVP model~\eqref{eq:NEVP} cannot (and does not, in our experience) properly capture the timescales of stress relaxation of these materials. Hence, we use a TEVP model, one of the simplest empirical models, to capture these observed behaviors \citep{Frigaard2019}.

A so-called structure parameter $\lambda(t)$ is used to describe the degree of solid-like behavior of a soft material \citep{Mewis2009}. Structural kinetics models of various complexity are used in the literature \citep{Jamali2017,DeSouzaMendes2011,Dullaert2006,Larson2019} to develop a governing equation for $\lambda(t)$.  We use a linear first-order kinetics model with buildup and breakage coefficients $k_{+}$ and $k_{-}$, respectively. The resulting TEVP model for the stress evolution \citep{Mahmoudabadbozchelou2021b,Saadat2023} takes the form:
\begin{subequations}
\begin{align}
    \frac{d\sigma}{dt} &= \frac{G}{\eta_s+\eta_p} \big\{ \sigma_y\lambda(t) + [\eta_s+\eta_p\lambda(t)] \dot \gamma(t) -\sigma(t) \big\},\label{eq:sigma_ode_TEVP}\\
    \frac{d\lambda}{dt} &= k_{+} [1-\lambda(t)] - k_{-}\dot{\gamma}(t)\lambda(t),\label{eq:lambda_ode}
\end{align}
\label{eq:TEVP-dim}%
\end{subequations}
where $\eta_p$ is the plastic viscosity, and all other quantities are as defined for Eq.~\eqref{eq:NEVP}. Generally, $\lambda$ is not measurable using a rheometer; $\lambda=1$ represents the fully structured (solid-like) state of the material, while $\lambda=0$ represents the fully unstructured (fluid-like) state of the material. In this model, the unknown parameters whose uncertainty is to be calibrated are $G$, $\eta_s$, $k_{+}$, $k_{-}$, $\sigma_y$, $\eta_p$.

Again, following on the observation of \citet{Zhang2025Data-drivenConduction}, we find an analytical solution to Eqs.~\eqref{eq:TEVP-dim} using the Laplace transform. Applying $\mathcal{L}$ to the ODEs in Eqs.~\eqref{eq:TEVP-dim}, we now find
\begin{widetext}
\begin{subequations}
\begin{align}
    s\bar{\sigma}(s) - \sigma(0) &= \frac{G}{\eta_s + \eta_p} \left[ -\bar{\sigma}(s) + \sigma_y \bar{\lambda}(s) + \eta_s \bar{\dot{\gamma}}(s) + \eta_p \mathcal{L}\{\lambda(t) \dot{\gamma}(t) \} \right], \label{eq:TEVP-dim-Laplace-sigma}\\
    s\bar{\lambda}(s) - \lambda(0) &= k_{+} \left[\frac{1}{s} - \bar{\lambda}(s) \right] - k_{-} \mathcal{L}\{\lambda(t) \dot{\gamma}(t) \}. \label{eq:TEVP-dim-Laplace-lambda}
\end{align}
\end{subequations}
For the special case of sudden step loading (startup rheometric protocol),  $\mathcal{L}\{\lambda(t) \dot{\gamma}(t) \} = \dot{\gamma}_0 \bar{\lambda}(s)$, which allows us to solve Eq.~\eqref{eq:TEVP-dim-Laplace-lambda} for $\bar{\lambda}$:
\begin{subequations}
\begin{equation}
    \bar{\lambda}(s) = \frac{k_{+}/s + \lambda(0)}{s + k_{+} + k_{-} \dot{\gamma}_0}.
    \label{eq:TEVP_lambda_s}
\end{equation}
Substituting Eq.~\eqref{eq:TEVP_lambda_s} into Eq.~\eqref{eq:TEVP-dim-Laplace-sigma}, we can solve for $\bar{\sigma}(s)$:
\begin{equation}
    \bar{\sigma}(s) 
    =  \frac{\sigma(0)}{s + \frac{G}{\eta_s + \eta_p}} + \frac{1}{s + \frac{G}{\eta_s + \eta_p}} \frac{\frac{G}{\eta_s + \eta_p} }{s}\Bigg[ \sigma_y \underbrace{\frac{k_{+} + s \lambda(0)}{s + k_{+} + k_{-} \dot{\gamma}_0}}_{s\bar{\lambda}(s)} + \; \eta_s \dot{\gamma}_0 +  \eta_p \dot{\gamma}_0 \frac{k_{+} + s \lambda(0)}{s + k_{+} + k_{-} \dot{\gamma}_0} \Bigg] .
    \label{eq:TEVP_sigma_s}
\end{equation}\label{eq:TEVP_sigma_lambda_s}%
\end{subequations}
Using tables of inverses and after some significant algebraic effort, Eqs.~\eqref{eq:TEVP_sigma_lambda_s} can be brought back to the time domain:
\begin{subequations}
\begin{align}
&\begin{aligned}
    \sigma(t) = \eta_s \dot{\gamma}_0 + (\sigma_y +\eta_p \dot{\gamma}_0) \lambda_\mathrm{equi} &+ \left\{ \sigma(0)-\eta_s \dot{\gamma}_0 + (\sigma_y +\eta_p \dot{\gamma}_0) \left[\frac{\lambda(0) - \lambda_\mathrm{equi} \Te}{\Te-1}\right] \right\} \re^{-\frac{G}{\eta_s+\eta_p} t}\\
    &+ (\sigma_y +\eta_p \dot{\gamma}_0) \left[\frac{\lambda_\mathrm{equi} -\lambda(0)}{\Te-1}\right] \re^{-(k_{+} + k_{-} \dot{\gamma}_0) t},
\end{aligned}\label{eq:TEVP-sol-sigma-step-strain}\\
&\begin{aligned}
    \lambda(t) = \lambda_\mathrm{equi} + \left[ \lambda(0) - \lambda_\mathrm{equi} \right] e^{-(k_{+} + k_{-} \dot{\gamma}_0)t}.
    \label{eq:TEVP-sol-lambda-step-strain}
\end{aligned}
\end{align}\label{eq:TEVP-sol-step-strain}%
\end{subequations}
\end{widetext}
Observe that the TEVP stress evolution~\eqref{eq:TEVP-sol-sigma-step-strain} is a superposition of \emph{two} exponentials with different time constants, as there is now the possibility of both stress relaxation (or buildup) \emph{and} microstructure relaxation.

Through the analytical solution~\eqref{eq:TEVP-sol-step-strain}, we have identified two key dimensionless numbers:
\begin{subequations}
    \begin{align}
    \Te &= 
    \frac{(\eta_s + \eta_p)/G}{(k_{+} + k_{-} \dot{\gamma}_0)^{-1}},
    \label{eq:Te}\\
    \lambda_\mathrm{equi} &= \frac{k_{+}}{k_{+} + k_{-} \dot{\gamma}_0} = \frac{1}{1 + \Th}. \label{eq:Th}
\end{align}
\end{subequations}
First, $\Te$ is the \emph{thixoelastic number} of \citet{Ewoldt2017}, representing the ratio of the characteristic bulk viscoelastic (stress relaxation) time scale to the characteristic microstructure evolution (thixotropic) time scale. As $\Te$ is the ratio of the time scales of the two exponents in Eq.~\eqref{eq:TEVP-sol-sigma-step-strain}, $\Te$ determines the dominant stress relaxation mechanism (thixotropy or viscoelasticity). Second, $\lambda_\mathrm{equi}$ is the asymptotic (equilibrium) value of $\lambda$ set via the competition of microstructure build and microstructure breakdown due to the imposed shear \citep{Mujumdar2002}. As $\lambda_\mathrm{equi}\to0$, the microstructure cannot buildup and $\lambda(t)\to0$ as $t\to\infty$ for any $\lambda(0)$. One can define a \emph{thixotropy number} from $\lambda_\mathrm{equi}$ as $\Th = k_{-}\dot{\gamma}_0/k_{+}$ so that for $\Th \gg 1$ corresponds to a significant breakdown of the microstructure and $\lambda_\mathrm{equi}\ll1$, and vice versa \citep{Mujumdar2002,Larson2019}. For further discussion on dimensionless numbers related to thixotropy, see \citep{Larson2019,Jamali2022,Joshi2024}.

\section{Bayesian Inference}
\label{sec::BI}
Bayesian inference is a method to solve inverse problems probabilistically~\citep{Gelman2013}. The model parameters are treated as random variables whose probability density functions (PDFs) have to be determined given the rheological data. 
We use $\mathbf{X}$ to denote the shear stress data obtained from experiments, and $\mathbf{\Theta}$ to denote the set of unknown model parameters (i.e., $\mathbf{\Theta} = [G, \eta_s, \sigma_y, K, n]$ for the NEVP model and $\mathbf{\Theta} = [G, \eta_s, k_{+}, k_{-}, \sigma_y, \eta_p]$ for the TEVP model). Posing a Bayesian inverse problem consists of three steps.  First, a PDF of unknown model parameters is constructed, which is known as the prior distribution $p(\mathbf{\Theta})$. This distribution encodes the analyst's belief of the uncertainty in the model's parameters \emph{before} seeing the data. Second, a probabilistic model of the measurement process, known as the likelihood function, is defined. We denote this by $p(\mathbf{X}|\mathbf{\Theta})$, which specifies the probability of observing $\mathbf{X}$ given the model parameters $\mathbf{\Theta}$. Third, the posterior distributions $p(\mathbf{\Theta}|\mathbf{X})$ is calculated using Bayes' rule:
\begin{equation}
    \label{eq:Bayes_theorem}
    p(\mathbf{\Theta}|\mathbf{X})=\frac{p(\mathbf{X}|\mathbf{\Theta})p(\mathbf{\Theta})}{p(\mathbf{X})},
\end{equation}
where $p(\mathbf{X})$ is a normalization constant, which is usually not available in closed form. To construct the posterior distribution, we use Markov Chain Monte Carlo (MCMC) sampling \citep{Robert2004TheAlgorithm} via the open-source Python package PyMC3 \citep{Salvatier2016ProbabilisticPyMC3}.

\subsection{Likelihood function}
In this subsection, we present the likelihood function. To model the experimental data, we  use the additive noise model, i.e., 
\begin{equation}
\label{eqn::model}
    \sigma_\mathrm{exp} = \sigma_\mathrm{model} + \tau_\mathrm{exp},
\end{equation}
where the subscript ``model'' stands for model predictions, and $\tau_\mathrm{exp}$ is the measurement uncertainty inherent to the experiments. We assume that the ``models,'' the NEVP solution given in Eq.~\eqref{eq:NEVP-sol-step-strain} and TEVP solution given in Eq.~\eqref{eq:TEVP-sol-step-strain}, are ``exact'' and, therefore, we do not explicitly calibrate model errors. We justify this appraoch by noting that the monotonic stress evolution data from experiments (Fig.~\ref{fig::expt_data}) is well explained by the NEVP and TEVP models \citep{Nagrani2023Data-drivenGreases}. Hence, we choose not to introduce additional parameters related to model error, which would also need to be calibrated. Furthermore, as noted by \citet{Rinkens2023}, having such ``fast-to-evaluate'' expressions for the models (here, given by the Laplace transform solutions above) is critical for an efficient implementation of the Bayesian approach.

We model the experimental error for the stress buildup in the DOWSIL\textsuperscript{TM} TC-5550 grease using a zero mean and finite variance Laplace distribution, $\tau_{\rm{exp}}\sim \text{Laplace} (0, b)$, where $b$ is the scale parameter. We anticipate longer tails in the shear stress measurements in the stress buildup regime, and hence, we use the Laplace distribution. Assuming independent measurements, the likelihood for the NEVP model (for which Eq.~\eqref{eq:NEVP-sol-step-strain} is used to evaluate $\sigma_\mathrm{model}$) is written as:
\begin{equation}
    \label{eqn::likelihood_NEVP}
    \begin{aligned}
    p(\sigma_\mathrm{exp}|\dot\gamma_\mathrm{exp},t,\mathbf{\Theta},b) &= \prod_{i=1}^{M} \prod_{j=1}^{N} p(\sigma_{ij}|\dot\gamma_{i},t_{ij},\mathbf{\Theta}_i,b)\\
    &= \prod_{i=1}^{M} \prod_{j=1}^{N} \Laplace (\sigma_{ij}|\sigma_\mathrm{model},b),
    \end{aligned}
\end{equation}
where $M$ is the number of different shear rates tested for the stress buildup regime, and $N$ is the number of data points collected during each startup experiment at a given $\dot\gamma_0$. 

For the stress relaxation of the  DOWSIL\textsuperscript{TM} TC-5622 grease, we model the experimental error using a zero mean normal distribution as $\tau_{\rm{exp}}\sim \mathcal{N} (0, c^2)$, where $c$ is the standard deviation.
Then, the likelihood for the TEVP model (for which Eq.~\eqref{eq:TEVP-sol-step-strain} is used to evaluate $\sigma_\mathrm{model}$) is written as:
\begin{equation}
    \label{eqn::likelihood_TEVP}
    \begin{aligned}
    p(\sigma_\mathrm{exp}|\dot\gamma_\mathrm{exp},t_\mathrm{exp},\mathbf{\Theta}, \lambda,c^2) &= \prod_{i=1}^{M} \prod_{j=1}^{N} p(\sigma_{ij}|\dot\gamma_{i},t_{ij},\mathbf{\Theta}_i,\lambda_i,c^2)\\
    &= \prod_{i=1}^{M} \prod_{j=1}^{N} \mathcal{N} (\sigma_{ij}|\sigma_\mathrm{model},c^2).
    \end{aligned}
\end{equation}
Here, in addition to the model parameters $\mathbf{\Theta}$, $\lambda$ needs to be estimated. One must also note that the consecutive shear stress data points are correlated. However, when the experimental data is conditioned on a model that represents the material behavior, our assumption of independence of consecutive data points in Eq.~\eqref{eqn::likelihood_NEVP} and Eq.~ \eqref{eqn::likelihood_TEVP} holds true. Furthermore, we do not allow the noise to vary with time due to the absence of an expression (or intuition) regarding this variation, which would be required to incorporate this effect (time-varying noise level) in the inference problem.

\subsection{Priors and hyperpriors}
\label{sec:priors}
The prior distributions on the model parameters $\mathbf{\Theta}$ reflect the analyst's belief about their uncertainty \emph{before} any data are observed. 
To allow variation of model parameters inferred across different shear rates, we use a hierarchical approach. 

Consider a single model parameter $\theta\in\mathbf{\Theta}$. The prior distribution of $\theta$ is parameterized by hyperparameters (denoted by $\mu_\theta$ and $\zeta_\theta$)  which themselves are distributions, called hyperpriors. Hence, every $\theta \in \mathbf{\Theta}$ is allowed to vary across different shear rates, i.e., if we use five shear rates to infer the model parameters, we have five different distributions for $\theta$. This approach allows us to quantify uncertainties that might exist in $\theta$ across different shear rates of interest. However, the hyperprior distributions ($\mu_\theta$ and $\zeta_\theta$) are constant across different shear rates (i.e., a global distribution exists across shear rates). We then sample from the posterior of the hyperpriors and use these samples to obtain a global distribution of $\theta$ that is valid across different shear rates.

\begin{figure}
    \centering
    \begin{subfigure}[t]{\linewidth}
      \includegraphics[width=\linewidth]{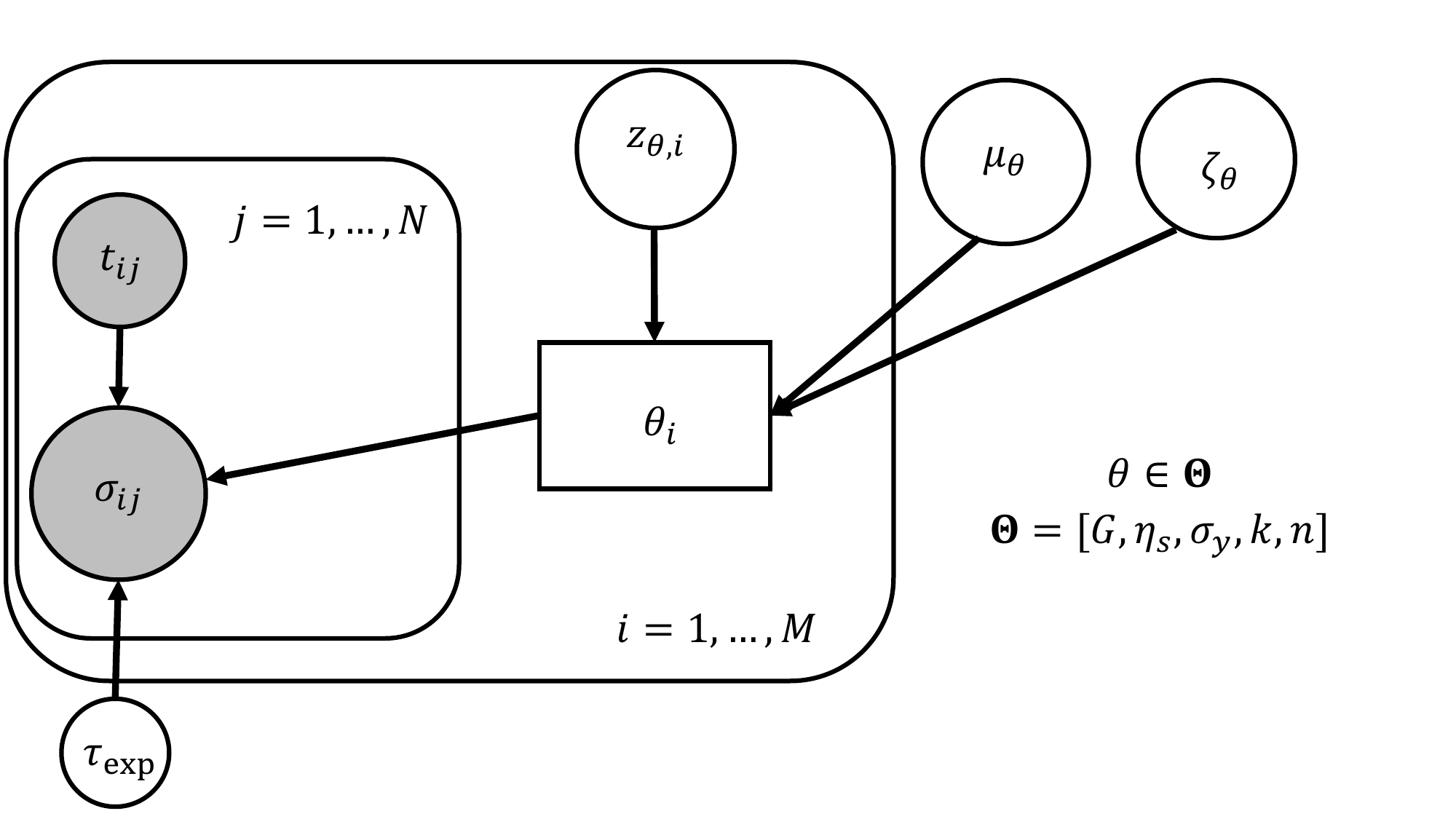}  
      \caption{}
      \label{fig:graohviz_nevp}
    \end{subfigure}
    \begin{subfigure}[t]{\linewidth}
      \includegraphics[width=\linewidth]{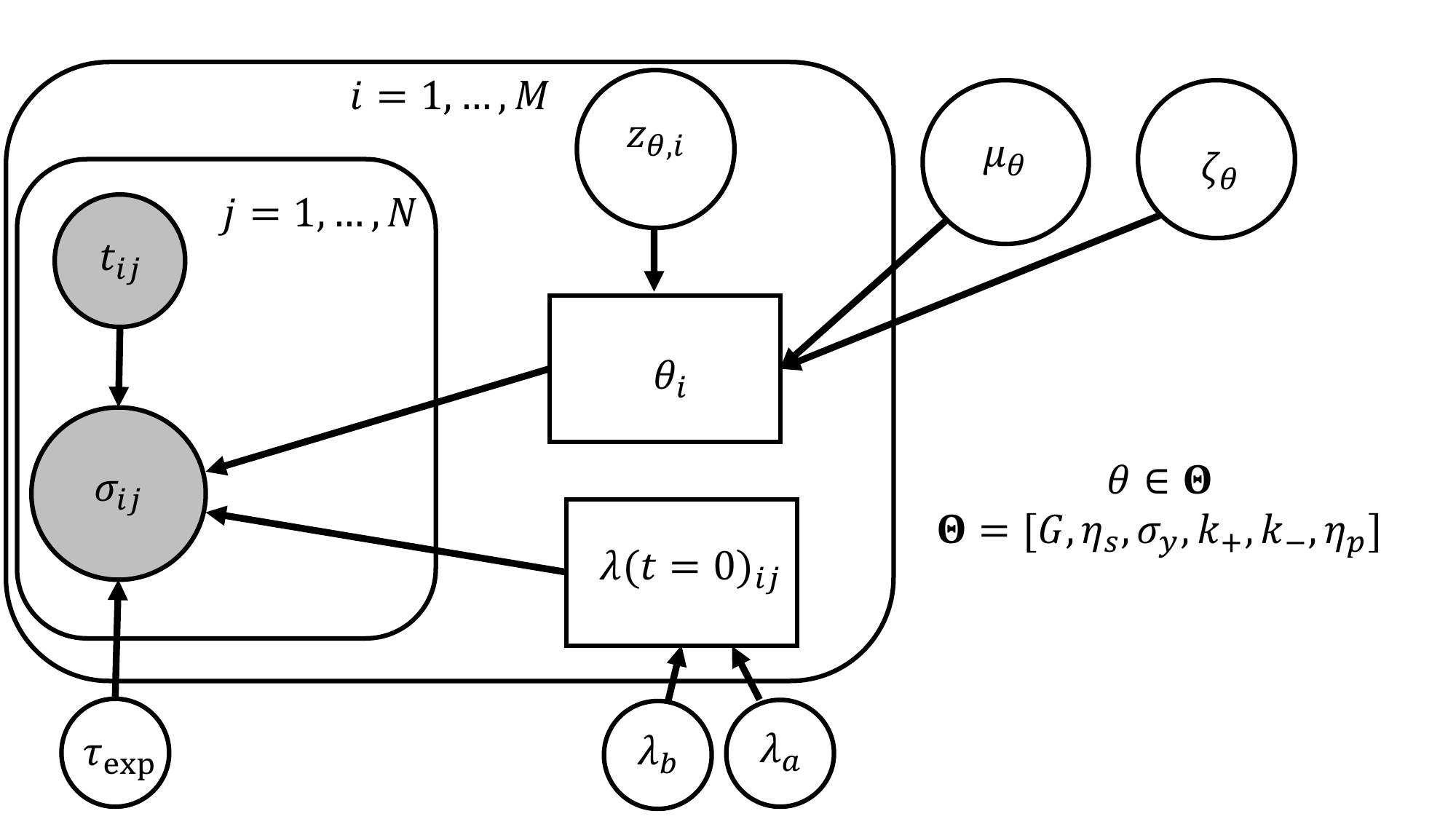}  
      \caption{}
      \label{fig:graphviz_tevp}
    \end{subfigure}
    \caption{Directed acyclic graphs of the proposed hierarchical Bayesian inference framework for (a) the NEVP model characterizing stress buildup and (b) the TEVP model characterizing stress relaxation. Nodes denoted by circles are random variables. Shaded circles are observed data from experiments. The square nodes represent deterministic variables.}
    \label{fig::graphviz_model}
\end{figure}

For each $\theta\in\mathbf{\Theta}$, for either the TEVP or the NEVP model, we use the following hyperpriors:
\begin{subequations}
    \begin{align}
    \label{eqn::hyperprior}
    \mu_\theta &\sim \Beta(2.0,2.0),\\
    \zeta_\theta &\sim \HalfNormal(0.05).
    \end{align}
\end{subequations}
The $\Beta$ distribution is used to ensure $\mu_{\theta}\in[0,1]$, while the $\HalfNormal$ distribution is used to ensure positive values for the hyperprior's standard deviations. Next, to enhance the stability of the sampling process, we use noncentered parametrization to define the priors. First, we define a standard normal distribution $\mathcal{N}$ for each model parameter as $z_\theta \sim \mathcal{N}(0,1)$. Second, we define the prior distributions as $\theta = \mu_\theta +  \zeta_\theta z_\theta$. Figure~\ref{fig::graphviz_model} shows the directed acyclic graph depicting the dependency between random variables. 

\begin{figure*}
    \centering
    \begin{subfigure}[t]{0.49\linewidth}
      \includegraphics[width=\linewidth]{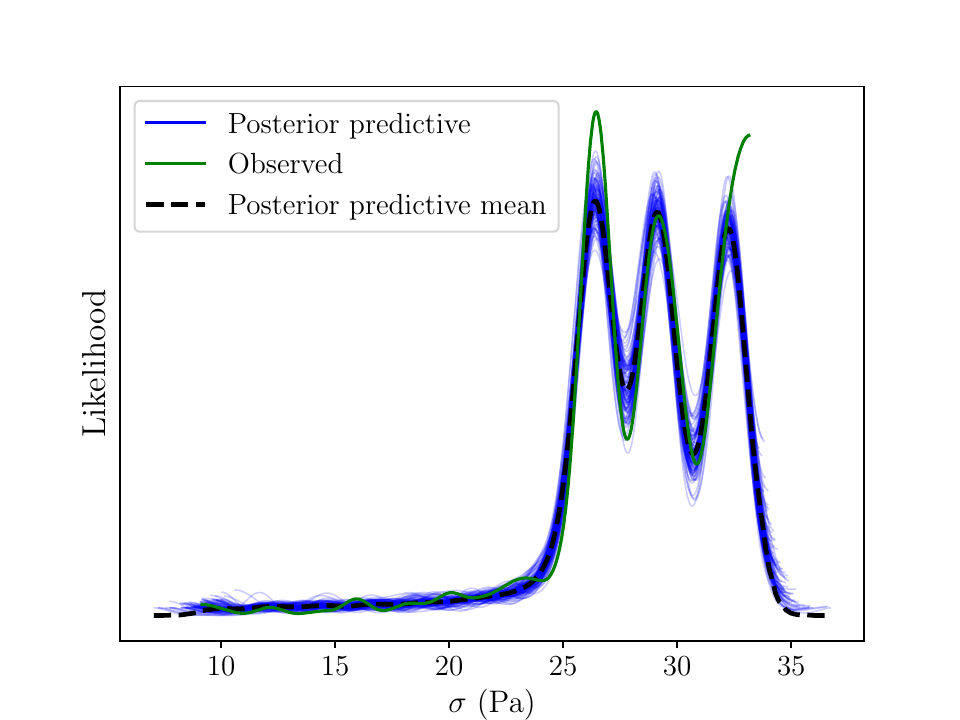}  
      \caption{}
      \label{fig:ppc_plot_nevp}
    \end{subfigure}
    \begin{subfigure}[t]{0.49\linewidth}
      \includegraphics[width=\linewidth]{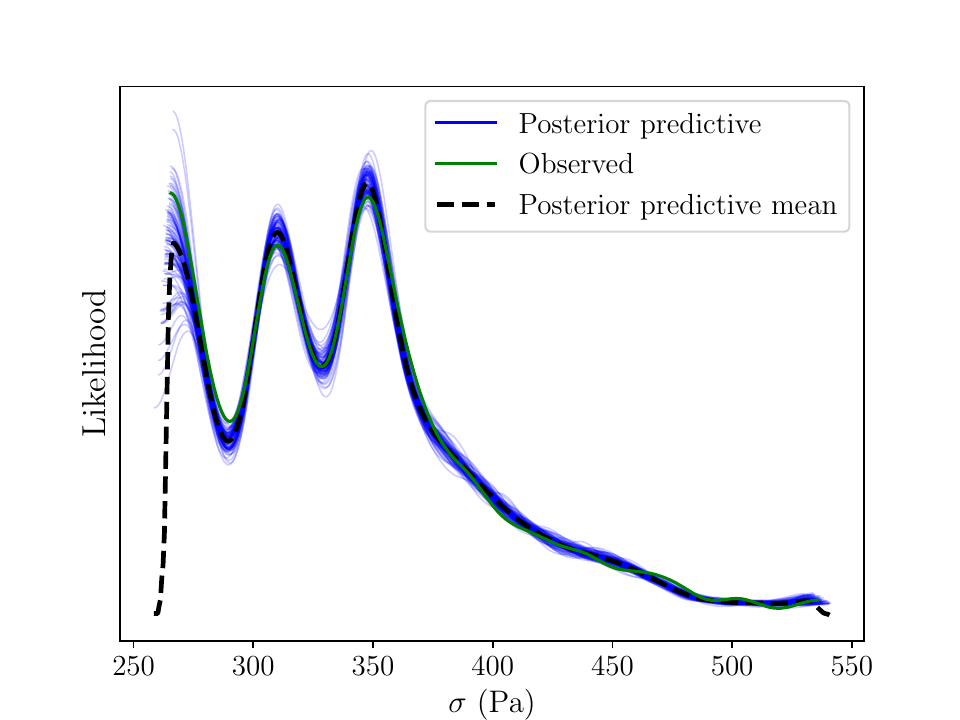}  
      \caption{}
      \label{fig:ppc_plot_tevp}
    \end{subfigure}
    \caption{Posterior predictive check (PPC) plots of the shear stress $\sigma$ for (a) the NEVP model and (b) the TEVP model comparing the models' ability to recreate the observational data, as identified in Fig.~\ref{fig::expt_data}.}
    \label{fig::ppc_plots}
\end{figure*}

The analytical solution~\eqref{eq:TEVP-sol-step-strain} of the TEVP model~\eqref{eq:TEVP-dim} is used to evaluate $\sigma_\mathrm{model}$ in the likelihood function in Eq.~\eqref{eqn::likelihood_TEVP}. However, Eq.~\eqref{eq:TEVP-sol-step-strain} depends on the initial state of structure parameter $\lambda(0)$, which is not measurable in rheometric experiments. Intuitively, we know that at higher shear rates, the microstructure ought to be more fluid-like (due to stronger shearing deformations. Following this intuition, we define a distribution to model the structure parameter's initial value: $\lambda(0)=\lambda_a \mathrm{exp} (-\lambda_b \dot\gamma)$. Here, $\lambda_a,\lambda_b \sim \Beta(2.0,5.0)$ are random variables whose posterior is inferred from the model to calculate $\lambda(0)$, which is then used to calculate the likelihood. Finally, exponential distributions are used as the prior distributions of the variances of the likelihood functions, i.e., $b \sim \ExpDist(2.0)$ and $c^2 \sim \ExpDist(2.0)$, where $2.0$ is the rate parameter of the exponential distribution.

\begin{figure*}
    \centering
    \begin{subfigure}[t]{0.4\linewidth}
      \includegraphics[width=\linewidth]{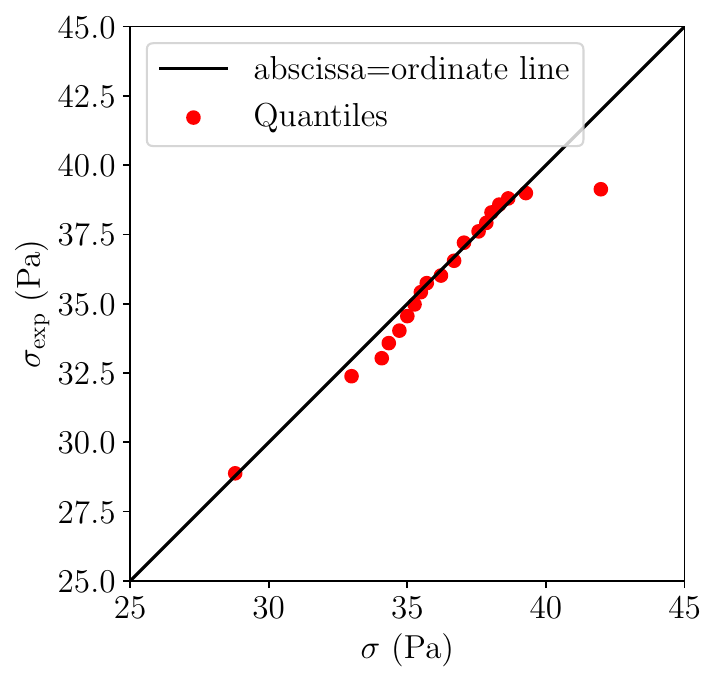}  
      \caption{}
      \label{fig:qq_plot_nevp}
    \end{subfigure}
    \qquad
    \begin{subfigure}[t]{0.4\linewidth}
      \includegraphics[width=\linewidth]{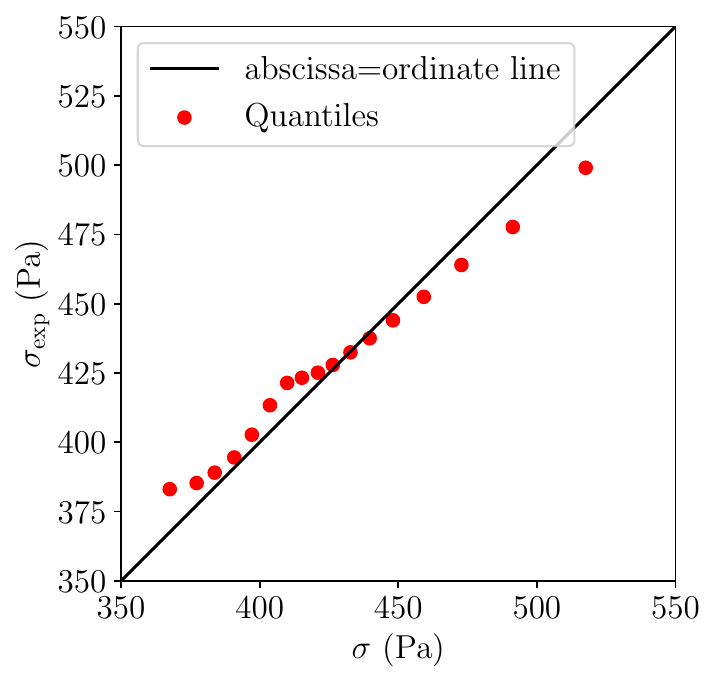}  
      \caption{}
      \label{fig:qq_plot_tevp}
    \end{subfigure}
    \caption{Quantile-quantile (Q-Q) plots of the shear stress $\sigma$ using $20$ quantiles for (a) the NEVP model and (b) the TEVP model comparing model predictions (with epistemic and aleatory uncertainties accounted for) and validation data, as identified in Fig.~\ref{fig::expt_data}.}
    \label{fig::qq_plots}
\end{figure*}

\subsection{Data-sampling methodology}
\label{sec:data_sampling}

We use the MCMC algorithm, specifically the No-U-Turn sampler \citep{Hoffman2014TheCarlo}, to sample the model parameters to obtain the posterior distribution. Next, to obtain the marginalized posteriors for any model parameter $\theta\in\mathbf{\Theta}$, we use the sum rule: 
\begin{equation}
    \label{eq::sum_rule}
    p(\theta|\sigma_\mathrm{exp}) = \int p(\theta,\mu_\theta,\zeta_\theta|\sigma_\mathrm{exp}) \,\rd\mu_\theta \rd\zeta_\theta .
\end{equation}
Using the product rule, $p(\theta,\mu_\theta,\zeta_\theta|\sigma_\mathrm{exp})=p(
\theta|\mu_\theta,\zeta_\theta)p(\mu_\theta,\zeta_\theta|\sigma_\mathrm{exp})$, where $p(\mu_\theta,\zeta_\theta|\sigma_\mathrm{exp})$ is sampled from the posterior distribution, and $p(\theta|\mu_\theta,\zeta_\theta)$ is obtained from the prior distribution with posterior-sampled $\mu_\theta$ and $\zeta_\theta$ as the inputs.  Note that $p(\mu_\theta,\zeta_\theta|\sigma_\mathrm{exp})$ is the marginalized posterior obtained from MCMC sampling. 

\begin{figure*}[ht]
    \centering
    \includegraphics[width=\linewidth]{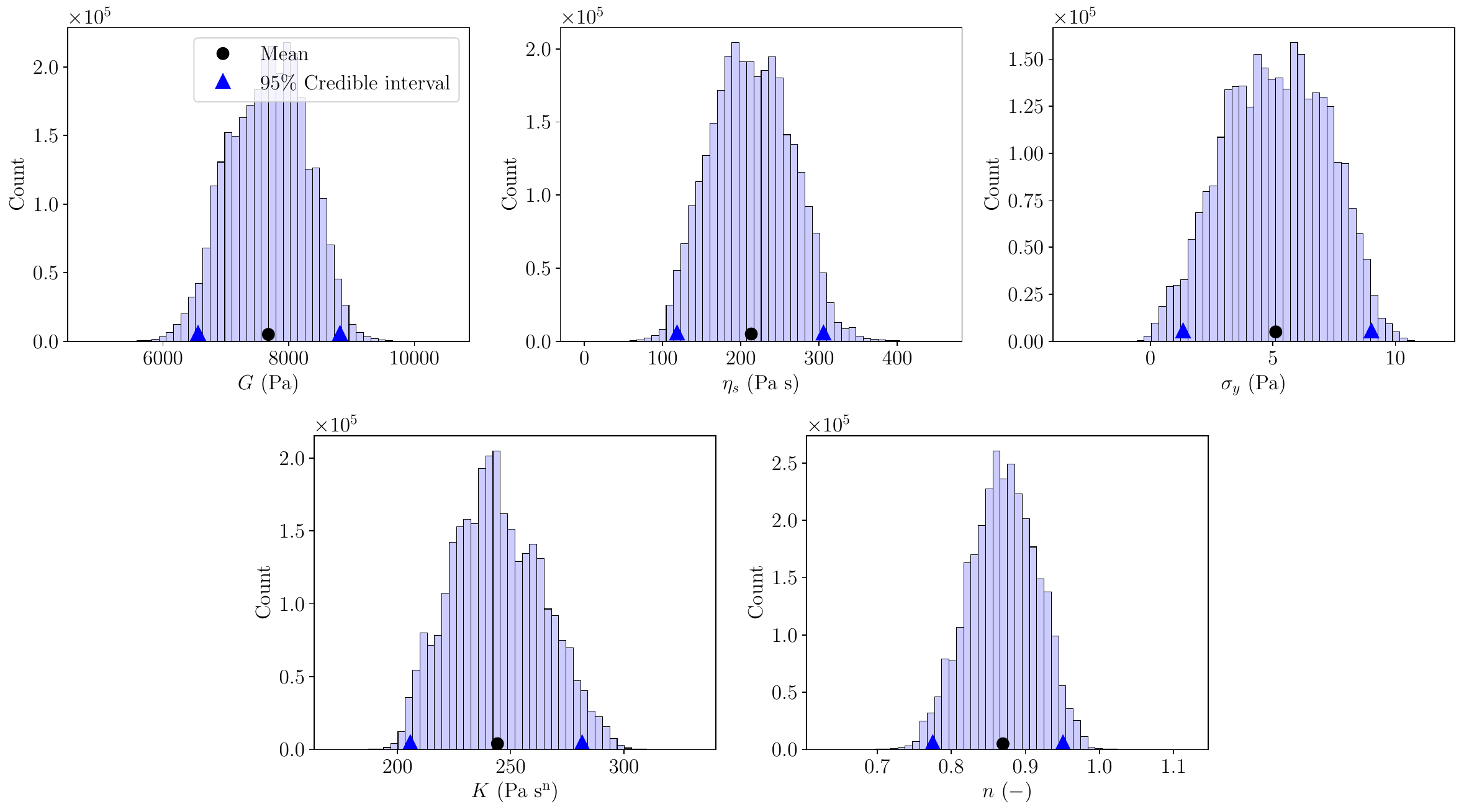}
    \caption{Sampled posterior distributions of the NEVP model's parameters, $G$, $\eta_s$, $\sigma_y$,  $K$, and $n$, describing the stress buildup in DOWSIL\textsuperscript{TM} TC-5550, when subjected to different step-strain shear rates $\dot\gamma_0$ at startup. The circles indicate the mean values and the triangles indicate the edges of the 95\% credible range.}
    \label{fig::NEVP_param_UQ}
\end{figure*}

\begin{figure*}[ht]
    \centering
    \includegraphics[width=\linewidth]{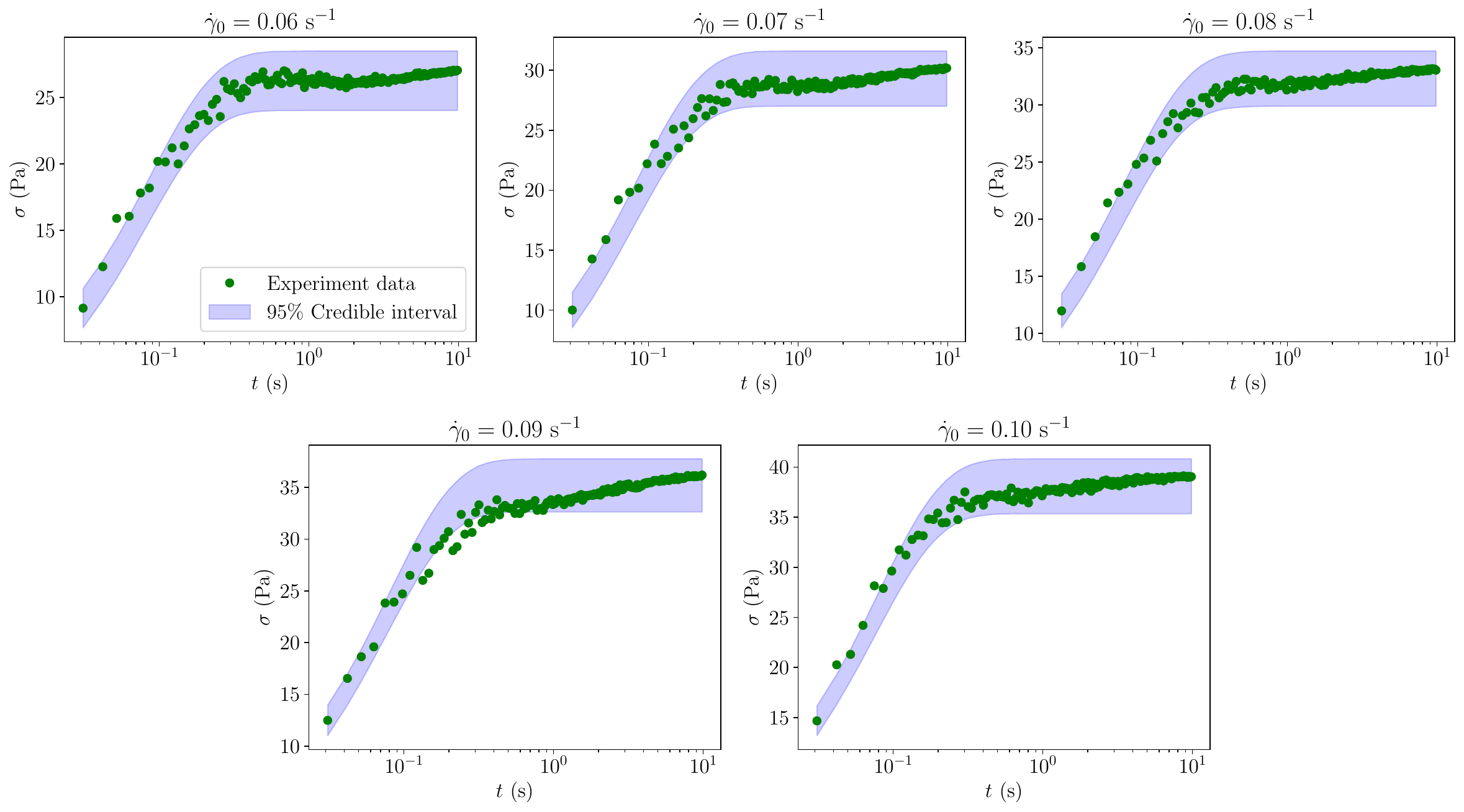}
    \caption{Uncertainty propagation of the NEVP model's parameters (Fig.~\ref{fig::NEVP_param_UQ}) into the transient shear stress profiles for DOWSIL\textsuperscript{TM} TC-5550. The panels' titles show the shear rate to which each plot corresponds.}
    \label{fig::NEVP_stress_UQ}
\end{figure*}

\section{Results and discussion}
\label{sec::RAndD}

In this section, we first verify the developed Bayesian inference framework (Sec.~\ref{sec::verification}). Then, we construct the marginalized posteriors for the NEVP and TEVP models' parameters (Sec.~\ref{sec:buildup} and \ref{sec:relax}, respectively). We also analyze these uncertainties' effect on the transient stress profiles. 

\begin{figure*}    
    \centering
    \includegraphics[width=\linewidth]{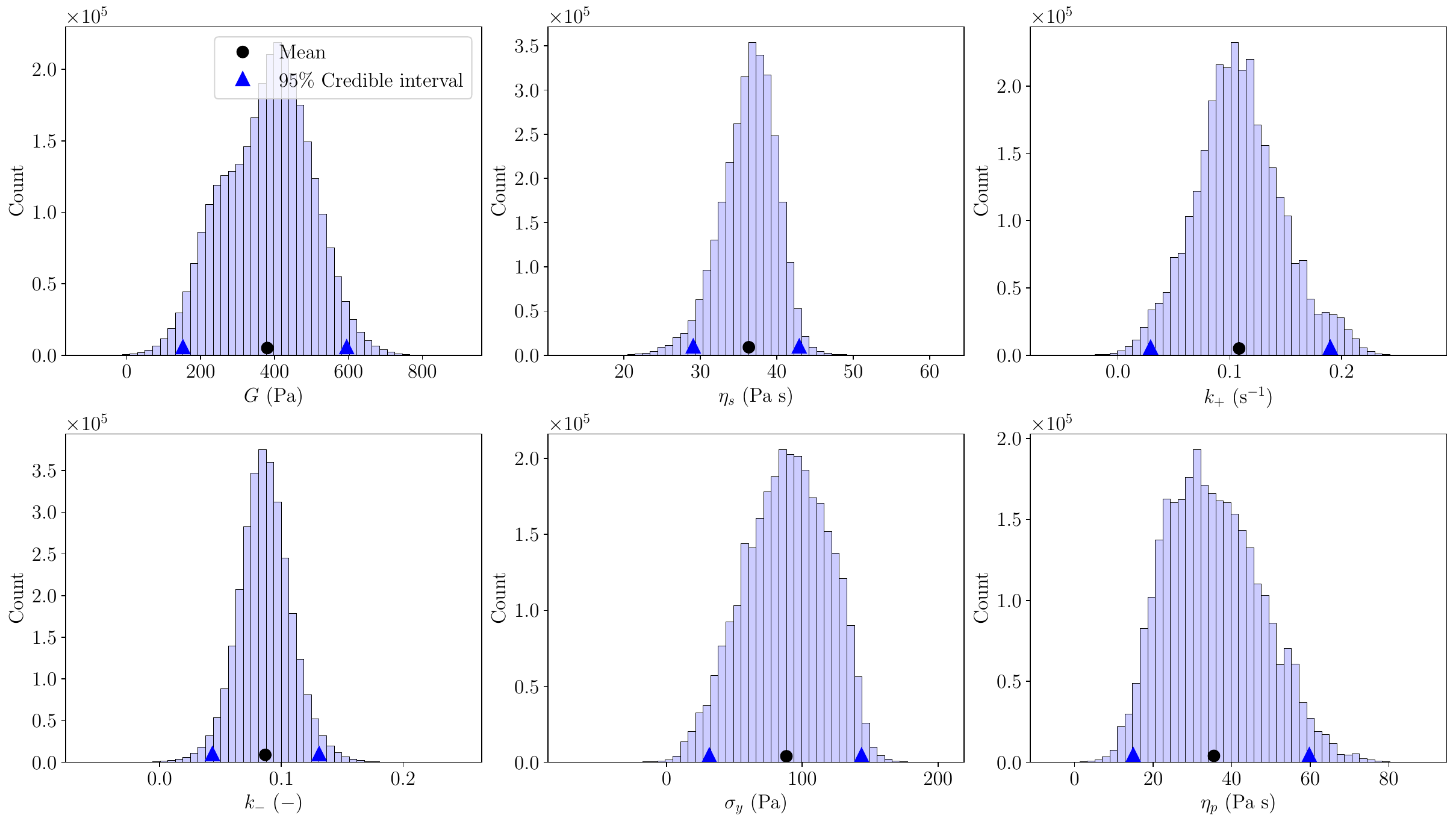}
    \caption{Sampled posterior distributions of TEVP model's parameters, $G$, $\eta_s$, $k_{+}$, $k_{-}$, $\sigma_y$, and $\eta_p$, describing the rheology of DOWSIL\textsuperscript{TM} TC-5622, when subjected to different step-strain shear rates $\dot\gamma_0$ at startup. The circles indicate the mean values and the triangles indicate the edges of the 95\% credible range. }
    \label{fig::TEVP_param_UQ}
\end{figure*}

\begin{figure*}
    \centering
    \includegraphics[width=\linewidth]{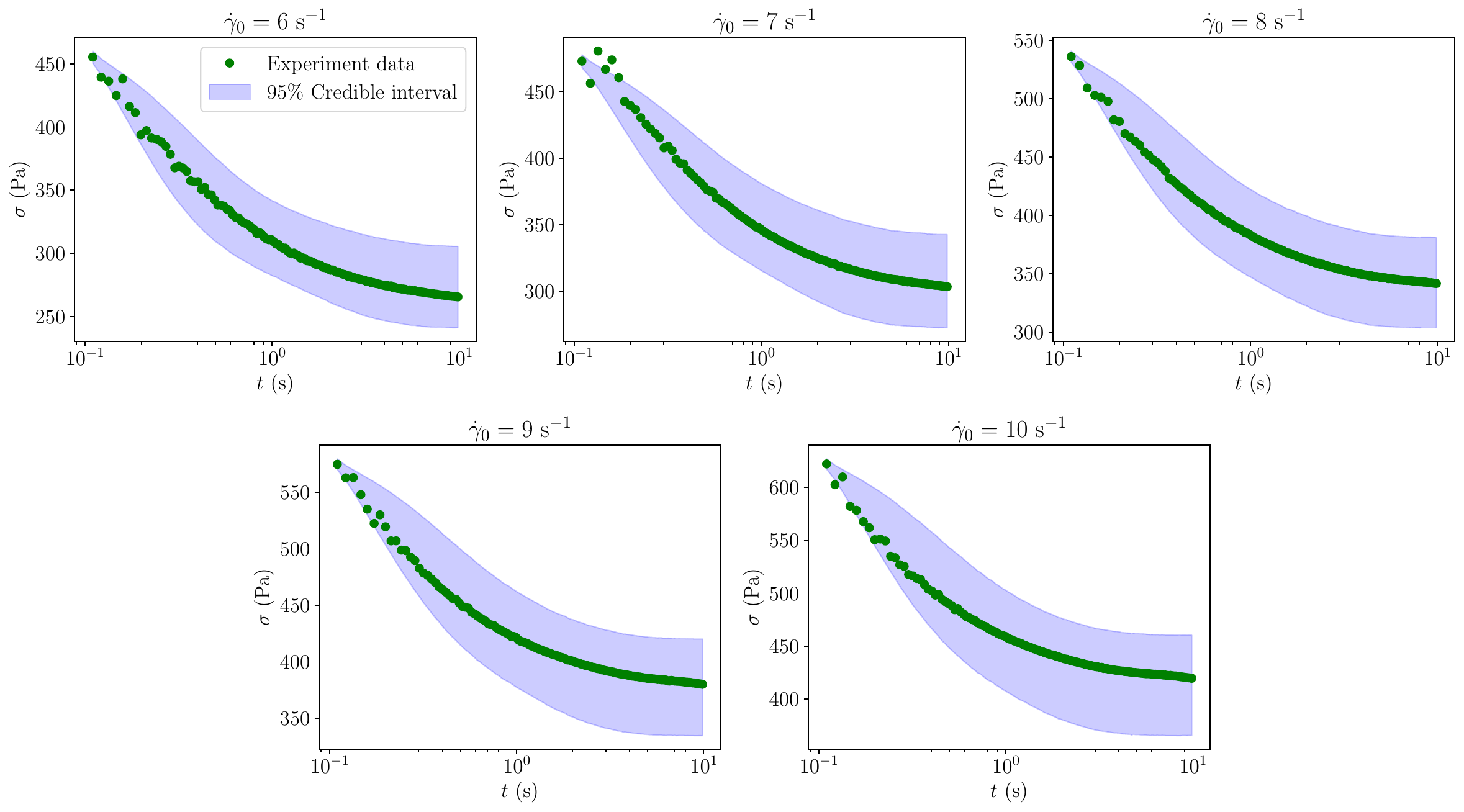}
    \caption{Uncertainty propagation of the TEVP model's parameters (Fig.~\ref{fig::TEVP_param_UQ}) into the transient shear stress profiles for DOWSIL\textsuperscript{TM} TC-5622. The panels' titles show the shear rate to which each plot corresponds.}
    \label{fig::TEVP_stress_UQ}
\end{figure*}

\subsection{Model verification}
\label{sec::verification}

Via MCMC sampling, we extracted $48000$ samples, out of which we discarded the first $8000$ to neglect the initial transients of the Markov chain. Next, we chose one in every $40$ samples to ensure they were not correlated.

In Fig.~\ref{fig::ppc_plots}, we verify the quality of posterior samples and the model performance by making posterior predictive check (PPC) plots. We sampled $100$ posterior predictions and compared them to the experimental observational data. The PPC plots demonstrate the TEVP and NEVP models' ability to reproduce the observational data, which was used to calibrate the models' parameters. The posterior predictions lie on top of the observed experimental data, thus providing a verification of the calibration of the NEVP and TEVP models to the experimental data. 

Next, in Fig.~\ref{fig::qq_plots}, we show quantile-quantile (Q-Q) plots. To generate model prediction data, we first sampled model parameters from their respective posterior distributions and used the sampled values to calculate the shear stress values via Eqs.~\eqref{eq:NEVP-sol-step-strain} and \eqref{eq:TEVP-sol-step-strain}. Next, we added the measurement error to the model predictions (obtained from the sampled values from the $\tau_\mathrm{exp}$ distribution) to capture the aleatory and epistemic uncertainties. Finally, we generated $20$ quantiles and compared the model predictions with uncertainties taken into account with the experimental validation data. The Q-Q plots thus validate the NEVP and TEVP models' performance by showing a strong correlation between the predicted and experimental shear stress values used as validation data. Specifically, the validation data are experiments at higher shear rates, which were not used in the calibration process. We observe that, for both rheological models, the calculated quantiles fall on top of the reference line, $\sigma_{\mathrm{exp}}=\sigma$, indicating accurate calibrations of the models' parameters and their uncertainties' propagation to the shear stress predictions.

However, a small deviation exists at higher stress values in Fig.~\ref{fig:qq_plot_nevp} and at lower stress values in Fig.~\ref{fig:qq_plot_tevp} due to the assumption of the NEVP and TEVP models being ``exact'' within the context of the inference problem. Naturally, a model bias (or error) exists that would need to be calibrated in addition to the calibration of the models' parameters. However, as explained earlier, we do not consider or calibrate the model errors because we are not focused on the model selection problem. Instead, we wish to understand the identifiability and calibration of the NEVP and TEVP models' parameters. To this end, we wish to ensure that the total number of parameters that need to be calibrated through the Bayesian framework is kept as low as possible (i.e., only the TEVP and NEVP models' parameters are calibrated). Due to the monotonic nature of shear stress profiles (recall Fig.~\ref{fig::expt_data}) and small number of outliers in the Q-Q plots (Fig.~\ref{fig::qq_plots}), we believe that the stress buildup and relaxation is accurately captured by the chosen rheological models, therefore model bias is not a key quantity that we need to calibrate in this study (although it may be for different models or different data).

\subsection{Uncertainty quantification of the stress buildup regime}
\label{sec:buildup}

The NEVP model~\eqref{eq:NEVP} is used to capture the stress buildup observed in experiments \citep{Nagrani2023Data-drivenGreases}. The posterior distributions of this model's inferred parameters ($G$, $\eta_s$, $\sigma_y$, $K$, and $n$) are shown in Fig.~\ref{fig::NEVP_param_UQ}. (For added visualization, in Appendix~\ref{app:prior_posterior}, we show the priors and posteriors on the same plot.) The mean and the corresponding $95\%$ credible intervals are labeled. Since there can be an infinite number of credible intervals that contain $95\%$ of the probability ``mass,'' we choose to show the highest posterior density interval (HDI). HDI is defined as the shortest interval that contains 95\% of the posterior mass. Importantly, all posterior distributions in Fig.~\ref{fig::NEVP_param_UQ} show a single mode, demonstrating the identifiability of each model parameter.

The uncertainties of the model's parameters give an estimate of the range of values that the rheological properties of DOWSIL\textsuperscript{TM} TC-5550 thermal grease can take, depending on which startup shear rates are used as observational data. Physically, we imagine that different rates of imposed shear rearrange the microstructure of the soft solid differently, which affects the bulk behavior and resistance to shear and, thus, the posterior of the inferred rheological parameters. Thus, it is of interest to determine how the shear rate in the experiments affects the inferred parameter values. To this end, we compare the HDI of each model parameter to the typical value estimated by solving the inverse problem using PINNs \citep{Nagrani2023Data-drivenGreases} in Table~\ref{tab::dow5550}. We observe that the parameter values previously inferred by PINNs mostly lie within the HDIs, which we learn can be quite wide. The values for $G$ and $\eta_s$ calibrated from PINNs fall outside the distribution predicted by the Bayesian approach, which could be due to the convergence at a local minima in training the PINN. However, note that this comparison primarily serves to verify the previously estimated values, not the inferred posterior distributions themselves.

Due to the uncertainty in the model's parameters, the predicted stress buildup response will also feature uncertainty, as shown in Fig.~\ref{fig::NEVP_stress_UQ}. To report the uncertainties of the transient shear stress profiles, we add the measurement noise to the model predictions and plot the $95\%$ credible intervals for the shear stress at each time to observe the stress evolution. The associated $95\%$ credible interval bands include uncertainties from epistemic and aleatory components arising due to limited data availability (at a given $\dot\gamma_0$) and random error inherent to the experiments, respectively. 

\begin{table}
    \caption{Comparison between high-density intervals (HDI) obtained from hierarchical Bayesian inference and the values estimated by PINNs \citep{Nagrani2023Data-drivenGreases} for the NEVP rheological model's parameters for DOWSIL\textsuperscript{TM} TC-5550.}
    \label{tab::dow5550}
    \centering
    \setlength{\tabcolsep}{5pt}
    \begin{tabular}{lll}
      \hline
      \hline
      Parameter & HDI & \cite{Nagrani2023Data-drivenGreases}  \\
      \hline
      $G$ (\si{\pascal}) & $[6558.2,8817.0]$ & $8972.0$  \\
      $\eta_s$ (\si{\pascal\second}) & $[118.5,305.6]$  & $363.0$ \\
      $\sigma_y$ (\si{\pascal}) & $[1.3,9.0]$ & $6.9$  \\
      $K$ (\si{\pascal \second \tothe{n}}) & $[205.7,281.6]$ & $214.2$ \\
      $n$ (--) & $[0.77,0.95]$ & $0.85$ \\
      \hline
      \hline      
    \end{tabular}
\end{table} 

Recall that the observational data at shear stress of $\dot\gamma_0 = 0.06~\si{\per\second}$, $0.07~\si{\per\second}$, and $0.08~\si{\per\second}$ is used to calibrate the model parameters. Hence, the Bayesian framework is expected to predict the uncertainty of shear stress profiles accurately for the observational data. More importantly, we also see good uncertainty predictions for the validation data (shear stress profiles for $\dot\gamma_0 = 0.09~\si{\per\second}$ and $0.1~\si{\per\second}$) in Fig.~\ref{fig::NEVP_stress_UQ}. Therefore, the developed hierarchical Bayesian inference framework can predict the uncertainties at higher shear rates in the stress buildup regime, suggesting that the model is generalizable throughout this regime.

To understand the impact of observational data on uncertainty calibration, in Appendix~\ref{app::shear_rates}, we recalibrated the NEVP model parameters using $\dot\gamma_0 = 0.08~\si{\per\second}$, $0.09~\si{\per\second}$, and $0.1~\si{\per\second}$ as observational data and compared the obtained distributions to the uncertainty calibrated using $\dot\gamma_0 = 0.06~\si{\per\second}$, $0.07~\si{\per\second}$, and $0.08~\si{\per\second}$ as observational data. We find that the uncertainty of the NEVP model's parameters and its predicted shear stress distributions are rather insensitive to the shear rates used for the calibration process (similar spread in the posterior distributions). Hence, any $\dot\gamma_0$ in the range of $0.06~\si{\per\second}$ to $0.1~\si{\per\second}$ can be used to for calibration without significantly affecting the uncertainty of the results.

\subsection{Uncertainty quantification of the stress relaxation regime}
\label{sec:relax}

The TEVP model~\eqref{eq:TEVP-dim} captures the stress relaxation observed in experiments \citep{Nagrani2023Data-drivenGreases}. The posterior distribution of this model's parameters ($G$, $\eta_s$, $\eta_p$, $\sigma_y$, $k_{+}$, and $k_{-}$), along with the means and the HDIs ($95\%$ credible intervals) are shown in Fig.~\ref{fig::TEVP_param_UQ}. (For added visualization, in Appendix~\ref{app:prior_posterior}, we show the priors and posteriors on the same plot.) The posterior distributions show a single mode for each model parameter, which indicates the identifiability of this linear TEVP model from the observational experimental data. 

In Table~\ref{tab::dow5622}, we compare the HDIs of model parameters obtained via the present hierarchical Bayesian inference to the typical values reported in the previous study \citep{Nagrani2023Data-drivenGreases}. It is seen that the values previously estimated by PINNs lie within the HDIs. However, from this Bayesian analysis, we learn that, based on the available data, these parameters can have significant uncertainty, especially the shear modulus $G$.

Additionally, recall that the Bayesian inference framework also obtains the initial value $\lambda(0)$ of the structure parameter needed to evaluate the exact solution~\eqref{eq:TEVP-sol-step-strain} to the TEVP model. Having assumed this value to be uncertain (recall Sec.~\ref{sec:priors}), we have also constructed the posterior distribution of $\lambda(0)$, from which the uncertainties could also be propagated onto $\lambda(t)$ via Eq.~\eqref{eq:TEVP-sol-lambda-step-strain}. However,  since there is no rheological data available for $\lambda(t)$, we cannot make any comparisons.

\begin{table}
     \caption{Comparison between high-density intervals obtained from hierarchical Bayesian inference and the values estimated by PINNs \citep{Nagrani2023Data-drivenGreases} for the TEVP rheological model's parameters for DOWSIL\textsuperscript{TM} TC-5622.}
     \setlength{\tabcolsep}{5pt}
    \begin{tabular}{lll}
      \hline
      \hline
      Parameter & HDI & \cite{Nagrani2023Data-drivenGreases} \\
      \hline
      $G$ (\si{\pascal}) & $[152.2,595.2]$ & $196.1$  \\
      $\eta_s$ (\si{\pascal\second}) & $[29.1,42.9]$ & $39.4$  \\
      $k_{+}$ (\si{\per \second}) & $[0.03,0.19]$ & $0.06$ \\
      $k_{-}$ (--) & $[0.04,0.13]$ & $0.06$  \\
      $\sigma_y$ (\si{\pascal}) & $[31.5,143.7]$ & $31.1$ \\
      $\eta_p$ (\si{\pascal\second}) & $[14.9,59.8]$ & $29.1$  \\
      \hline
      \hline
    \end{tabular}
   
    \label{tab::dow5622}
\end{table} 

Next, the uncertainties of the model's parameters are propagated to establish the HDIs for the stress relaxation profiles, shown in Fig.~\ref{fig::TEVP_stress_UQ}. The methodology to propagate the uncertainties is the same as in Sec.~\ref{sec:buildup}. That is, we add the experimental measurement error to the model predictions at each time point and then calculate the $95\%$ credible intervals. The shear stress measurements at $\dot\gamma_0 =6~\si{\per\second}$, $7~\si{\per\second}$, and $8~\si{\per\second}$ are used for observational data, and the model predictions are evaluated at $\dot\gamma_0=9~\si{\per\second}$ and $10~\si{\per\second}$. 

Figure~\ref{fig::TEVP_stress_UQ} shows that the developed hierarchical Bayesian framework can quantify the uncertainties of the observational and validation data sets. The epistemic uncertainty arises from the data being available at only a few shear rates ($\dot\gamma_0 = 6~\si{\per\second}$ to $10~\si{\per\second}$) in the stress relaxation regime. Meanwhile, the aleatory uncertainty is due to random error, such as due to shearing history effects on the sample used for the experiments. 

To understand the impact of observational data on uncertainty calibration, in Appendix~\ref{app::shear_rates}, we recalibrated the TEVP model parameters using stress profiles at $\dot\gamma_0 = 8~\si{\per\second}$, $9~\si{\per\second}$, and $10~\si{\per\second}$ as observational data and compared the obtained distributions to the uncertainty calibrated using stress profiles at $\dot\gamma_0 = 6~\si{\per\second}$, $7~\si{\per\second}$, and $8~\si{\per\second}$ as observational data. The general takeaway is that the posterior distributions of the TEVP model parameters have a narrower spread when calibrated at higher shear rates, suggesting that rheometric experiments at higher $\dot\gamma_0$ are more insightful for model calibration. 

\begin{figure}
    \centering
    \begin{subfigure}[t]{0.9\linewidth}
      \includegraphics[width=\linewidth]{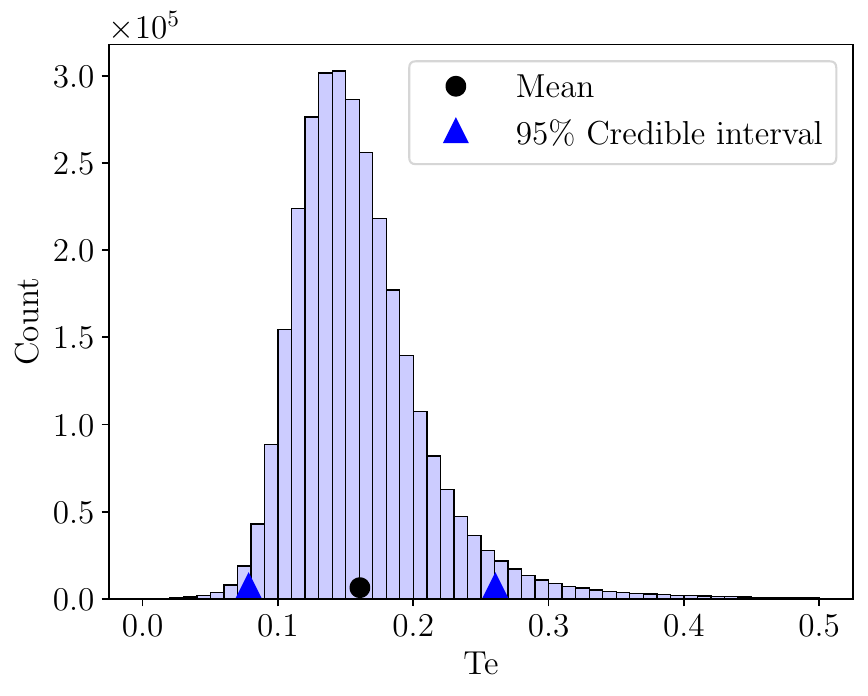}  
      \caption{}
      \label{fig:Te}
    \end{subfigure}
    \\
    \begin{subfigure}[t]{0.9\linewidth}
      \includegraphics[width=\linewidth]{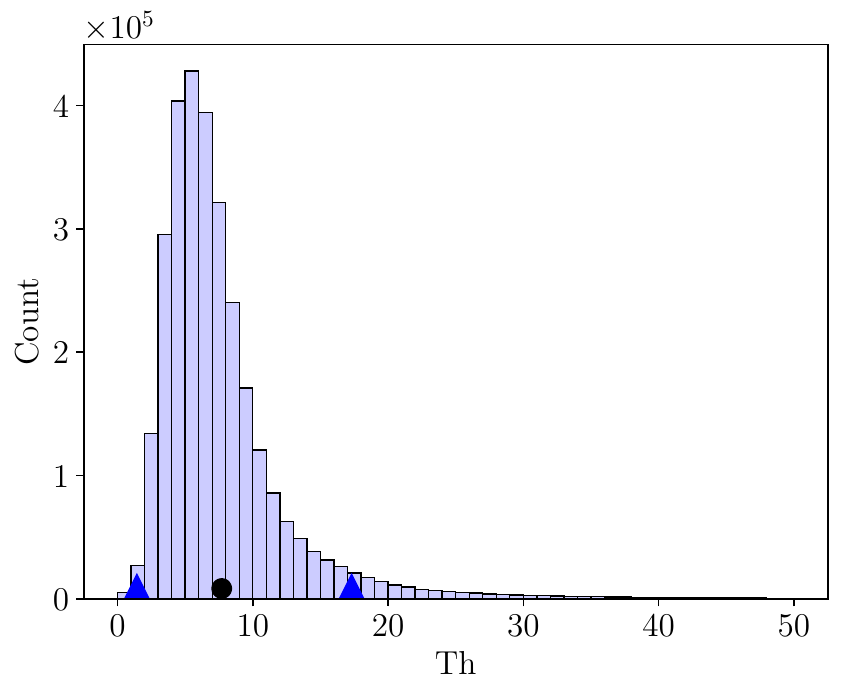}  
      \caption{}
      \label{fig:Th}
    \end{subfigure}
    \caption{Uncertainty propagation of the TEVP model's parameters into (a) the thixoelastic number, $\Te$, defined in Eq.~\eqref{eq:Te} and (b) the thixotropy number, $\Th$, defined below Eq.~\eqref{eq:Th}.}
    \label{fig::Thixotropy_nos}
\end{figure}

Finally, in Fig.~\ref{fig::Thixotropy_nos}, we propagate the posterior distributions of the TEVP model's parameters into the dimensionless numbers  $\Te$ (defined in Eq.~\eqref{eq:Te}) and $\Th$ (defined below Eq.~\eqref{eq:Th}). From the distribution of $\Te$ in Fig.~\ref{fig:Te}, we learn that  DOWSIL\textsuperscript{TM} TC-5622's stress response is dominated by thixotropic effects due to microstructure evolution, more so than its viscoelasticity.  This observation is also supported by the distribution of $\Th$ in Fig.~\ref{fig:Th}, wherein the mean value of $\Th\gg1$, again indicating strongly thixotropic stress response (and, thus, considerable microstructure breakdown) within the sheared material. 

\section{Conclusion}
\label{sec::Conclusion}
We introduced a hierarchical Bayesian framework to characterize uncertainties of the rheological parameters of two commercial thermal greases, which comprise of high thermal conductivity particles dispersed within a polymer matrix. Thermal greases are typically subjected to thermo-mechanical stresses within electronic packages, which can cause significant microstructural rearrangements within these materials, resulting not only in complex behaviors like thixotropy but also in uncertainties in the inferred bulk properties of interest, which are often reported \emph{without} any distribution densities in prior literature and product datasheets. 

Specifically, we reanalyzed startup shear stress data from previous experiments \citep{Nagrani2023Data-drivenGreases} on DOWSIL\textsuperscript{TM} TC-5550 and DOWSIL\textsuperscript{TM} TC-5622 thermal greases exhibiting stress buildup and relaxation, respectively, each of which is captured by a different rheological model. We calibrated these models' parameters using the developed hierarchical Bayesian inference framework, which allows the model parameters' distributions to vary across different startup shear rates at which experiments were conducted. For each parameter, the posterior distributions across shear rates were constructed and then pooled into a single distribution. This approach identified the effect of observational data on the uncertainty of the inferred parameter values. 

We observed a single mode in the posterior distribution of each rheological parameter, demonstrating the chosen models' identifiability from the available data. Finally, we propagated the calibrated uncertainty of the models' parameters to the transient shear stress profiles to account for epistemic and aleatoric uncertainty sources in the model predictions. The hierarchical Bayesian approach revealed the role of the shear rates at which startup experiments were conducted, which is not evident from inverse methods (such as those based on PINNs). Specifically, the uncertainty of the TEVP model's parameters (in the stress relaxation regime) is reduced (i.e., the posterior distributions are narrower) if the observational data is taken from the experiments at higher shear rates. 

Interestingly, through the exact model solutions we developed and used in the Bayesian inference, our approach does not require ``learning'' the microstructure evolution under the linear TEVP model, for which there is no experimental data to assimilate. Instead, the initial value of the microstructure variable can be treated as an uncertain calibration parameter within the Bayesian framework. Furthermore, despite having no data on the microstructure, the TEVP model's parameters can be used to construct the posterior distributions of the key dimensionless groups---a thixoelastic and thixotropy number. These distributions suggest that the stress response of DOWSIL\textsuperscript{TM} TC-5622 is strongly thixotropic, indicating significant microstructure breakdown during its stress relaxation.

An interesting avenue for future research would be to use the developed hierarchical Bayesian inference framework to compare the uncertainties of the rheological properties and, hence, the spread of thermal resistances of thermal greases by using rheological data obtained before \emph{and} after the degradation of the materials.  Understanding the uncertainties of the inferred thermorheological parameters can gauge the utility and predictive power of different models used to evaluate the performance of such materials for thermal applications. Finally, it would also be of interest to incorporate data from other types of rheological tests, e.g., large-amplitude oscillatory shear \citep{Kamani2023UnderstandingFluids,Agrawal2025FeaturesLAOS}, into the proposed hierarchical Bayesian inference methodology to understand the consequences of the choice of rheological tests on the model calibration, and determine whether the uncertainties of specific parameters can be reduced by using data from different types of experiments.


\section*{Author Contributions}
\textbf{Pranay P.\ Nagrani}: conceptualization (equal); data curation (lead); formal analysis (lead); investigation (equal); methodology (equal); validation (lead); visualization (lead); writing - original draft (lead).
\textbf{Akshay J. Thomas}: conceptualization (equal); formal analysis (supporting); investigation (equal);  writing - original draft (supporting); writing - review \& editing (lead).
\textbf{Amy M.\ Marconnet}: writing - review \& editing (equal); supervision (supporting); project administration (supporting).
\textbf{Ivan C.\ Christov}: conceptualization (equal); formal analysis (equal); investigation (supporting); methodology (supporting); writing - original draft (equal); writing - review \& editing (equal); supervision (lead); project administration (lead).

\section*{Acknowledgements}
P.P.N.\ was supported by a Bilsland Dissertation Fellowship from The Graduate School at Purdue University.

A.J.T.\ was supported by the AFOSR program on materials for extreme environments under grant number FA09950-22-1-0061.

I.C.C.\ acknowledges the donors of the American Chemical Society Petroleum Research Fund for partial support of his research under ACS PRF award \# 68986-ND9.

We thank Akash Mattupalli for discussions and independent testing of pieces of the present approach in its early stages.

Partial financial support for this work was provided by members of the Cooling Technologies Research Center, a graduated National Science Foundation Industry/University Cooperative Research Center at Purdue University, USA.


\section*{Data Availability}
The data that support the findings of this study are openly available in a repository archived at  \url{https://dx.doi.org/10.5281/zenodo.15616035}.


\appendix

\section{Sensitivity analysis of shear rates used as observational data}
\label{app::shear_rates}
In this appendix, we assess the influence of observational data used to calibrate the uncertainties of the model's parameters. Unlike in the main text, we now use the experiments at the higher shear rates as observational data. Then, we compare the predicted uncertainties of the shear stress response to the validation data from the experiments at the lower shear rates.

\subsection{Stress buildup regime}

\begin{figure*}
    \centering
    \includegraphics[width=\linewidth]{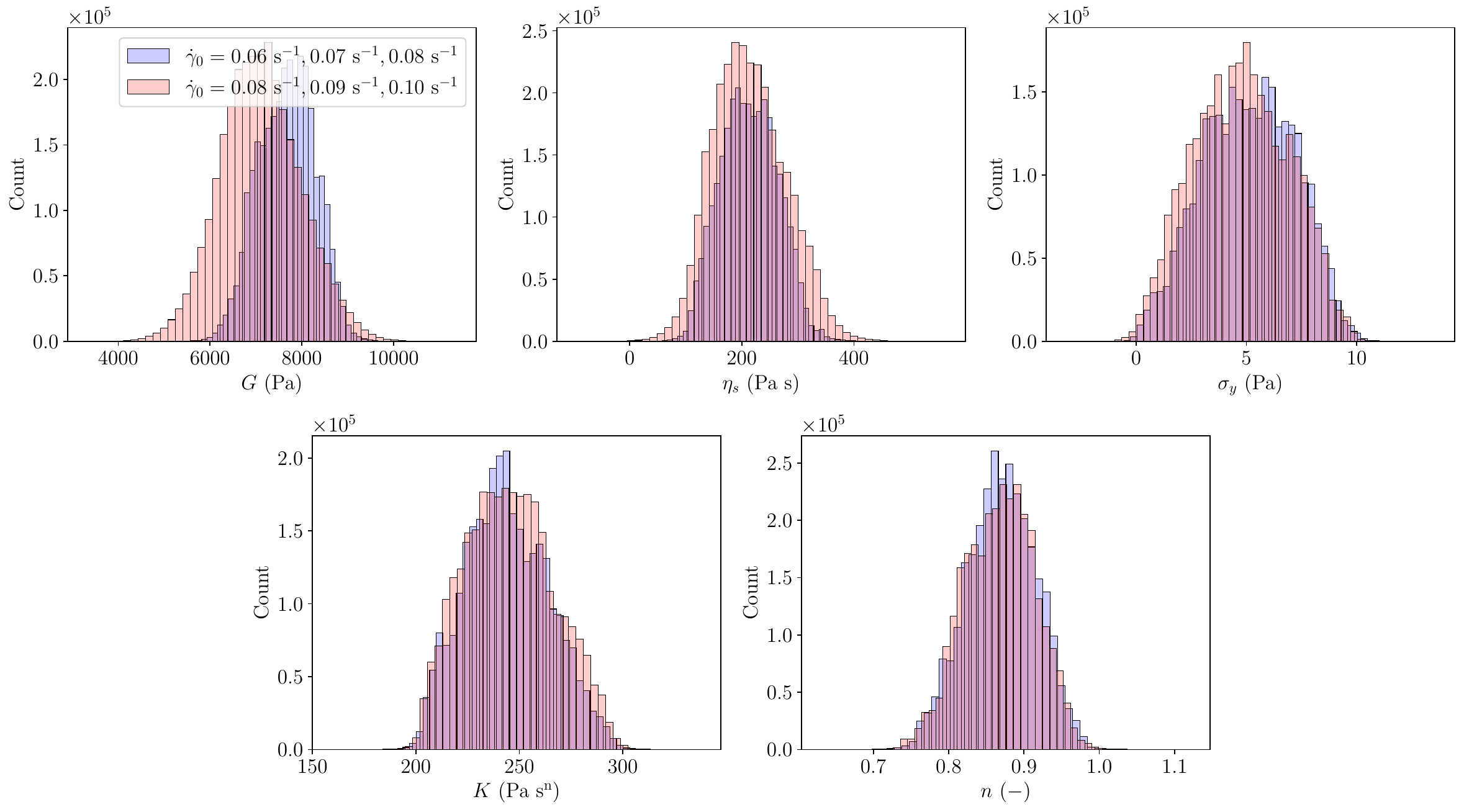}
    \caption{Influence of observational data used to perform the Bayesian inference (Sec.~\ref{sec:buildup}) for the NEVP model on the calibration of the uncertainty of the model's parameters $G$, $\eta_s$, $\sigma_y$,  $K$, and $n$.}
    \label{fig::NEVP_param_shear_rate}
\end{figure*}

In Fig.~\ref{fig::NEVP_param_shear_rate}, we compare the posterior distributions of the NEVP model's parameters obtained using observational data with $\dot\gamma_0 = 0.08~\si{\per\second}$, $0.09~\si{\per\second}$, and $0.1~\si{\per\second}$, to those obtained using observational data with $\dot\gamma_0 = 0.06~\si{\per\second}$, $0.07~\si{\per\second}$, and $0.08~\si{\per\second}$. The posterior distributions for the two different sets of observational shear rates fall mostly on top of each other for all model parameters except $G$, highlighting the sensitivity of the elastic shear modulus on the shear rates used for calibrating the model's parameters. 

\begin{figure*}
    \centering
    \includegraphics[width=\linewidth]{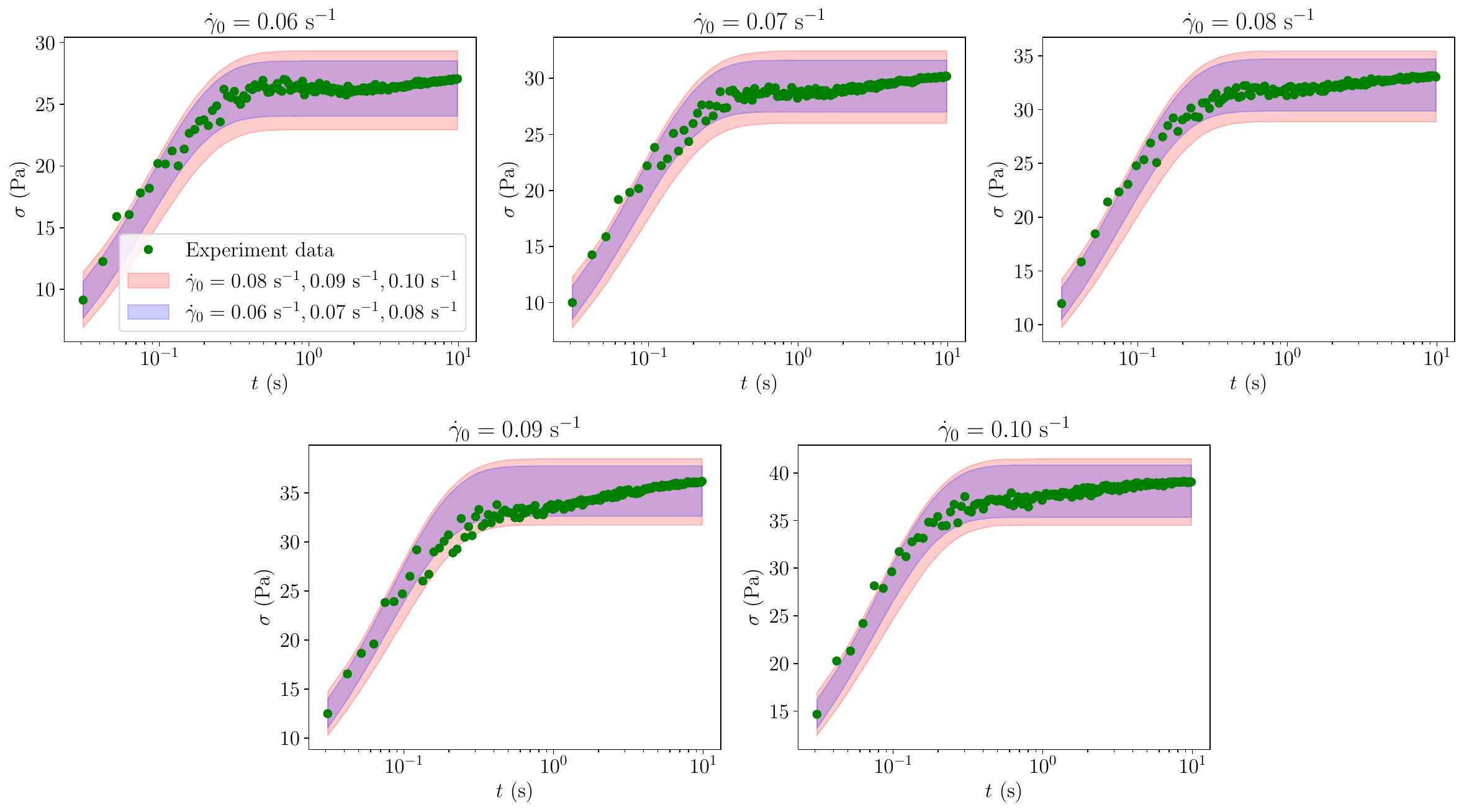}
    \caption{Propagated uncertainties of the NEVP model's parameters calibrated in Fig.~\ref{fig::NEVP_param_shear_rate} to the stress buildup seen in DOWSIL\textsuperscript{TM} TC-5550.}
    \label{fig::NEVP_shear_stress_shear_rate}
\end{figure*}

Next, in Fig.~\ref{fig::NEVP_shear_stress_shear_rate}, we compare the propagated uncertainty of the shear stress profiles for two sets of observational data. We see that the stress curves' HDIs more or less lie on top of each other across the entire range of time. Hence, we conclude that the stress buildup regime characterized by the NEVP model can be calibrated with observational data for \emph{any} shear rate in the range of $0.06~\si{\per\second}$ to  $0.1~\si{\per\second}$.

\begin{figure*}
    \centering
    \includegraphics[width=\linewidth]{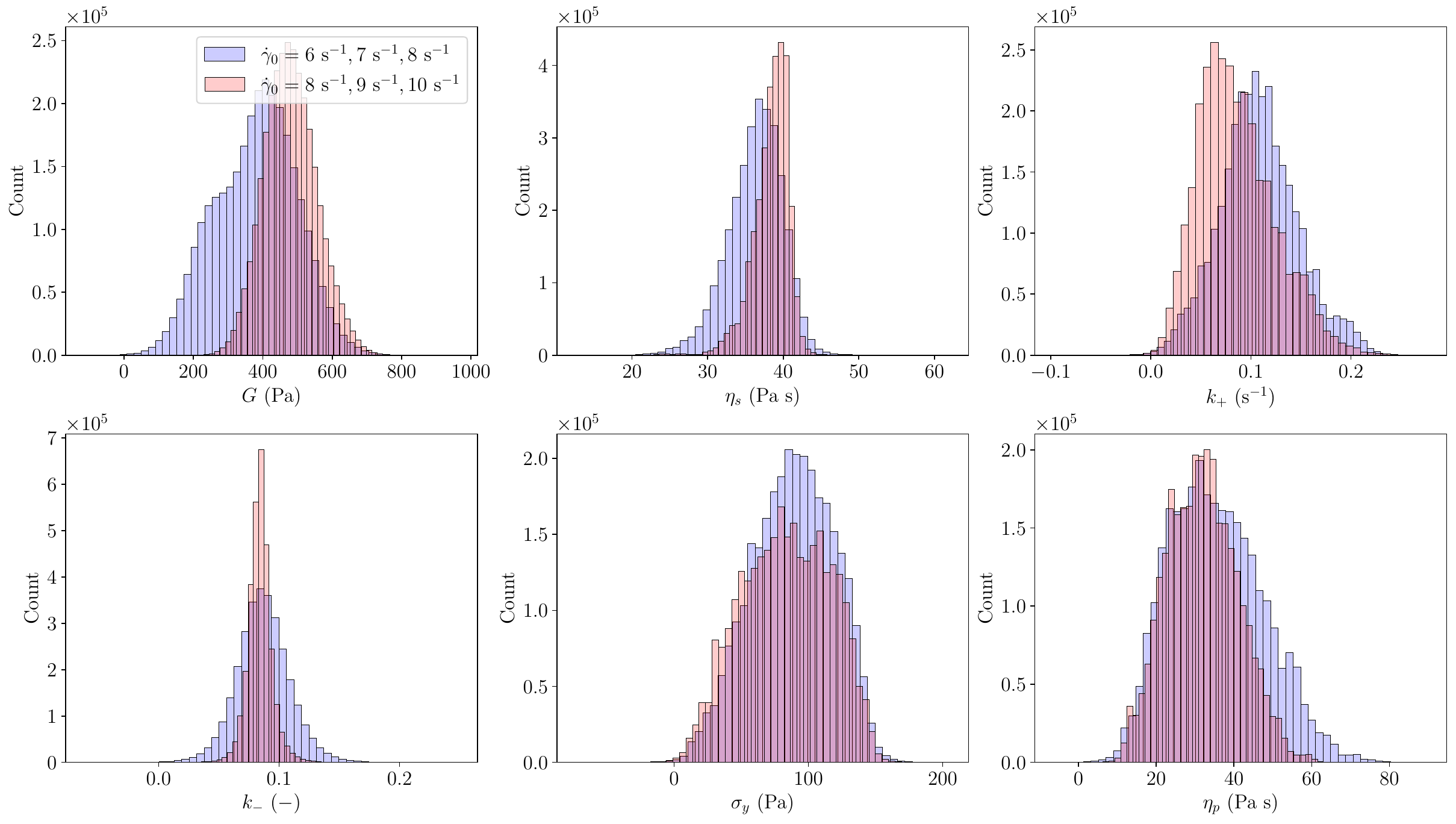}
    \caption{Influence of observational data used to perform the Bayesian inference (Sec.~\ref{sec:relax}) for the TEVP model  on the calibration of the uncertainty of the model's parameters $G$, $\eta_s$, $k_{+}$, $k_{-}$, $\sigma_y$, and $\eta_p$.}
    \label{fig::TEVP_param_shear_rate}
\end{figure*}

\begin{figure*}
    \centering
    \includegraphics[width=\linewidth]{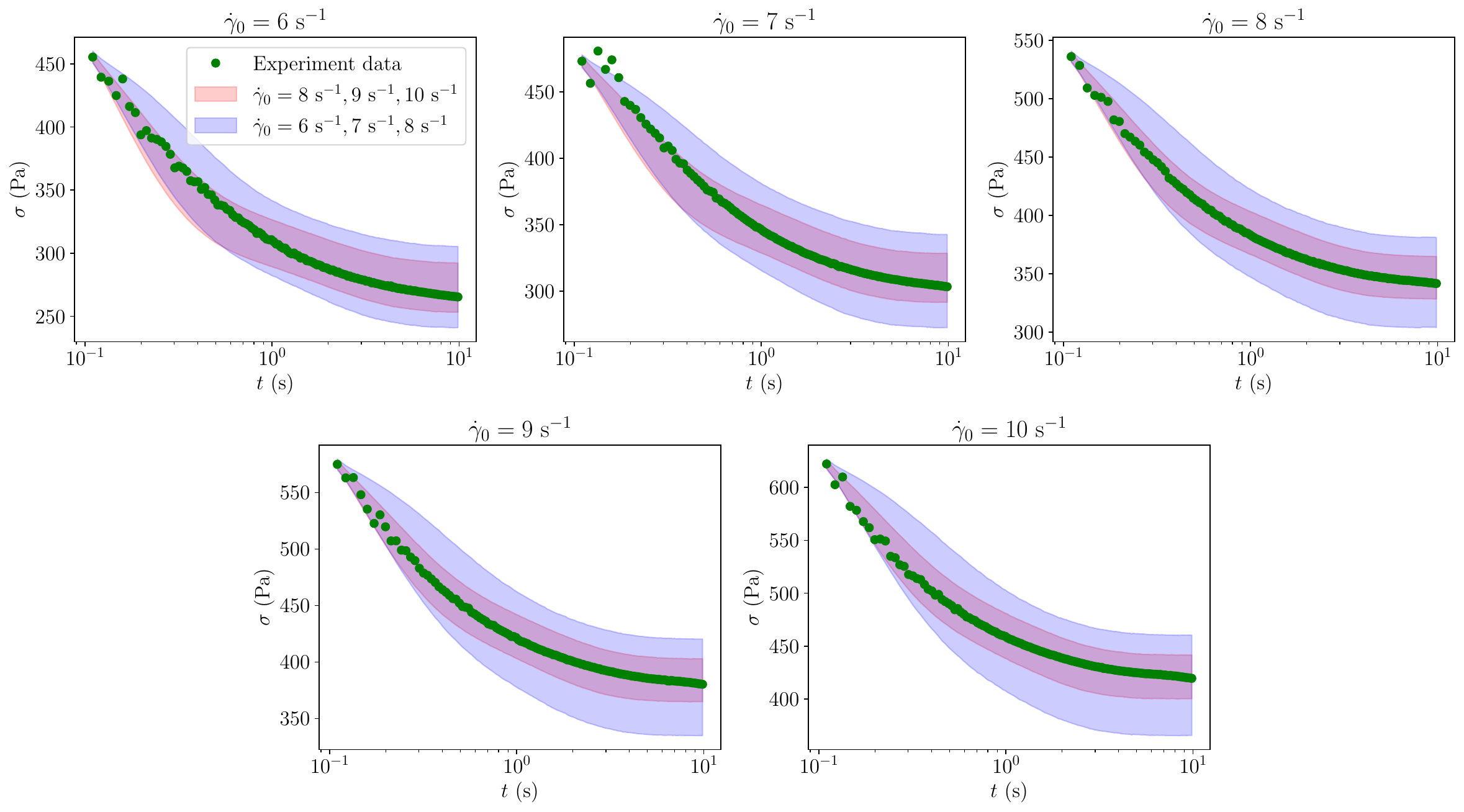}
    \caption{Propagated uncertainties of the TEVP model's parameters calibrated in Fig.~\ref{fig::TEVP_param_shear_rate} to the stress relaxation seen in DOWSIL\textsuperscript{TM} TC-5622.}
    \label{fig::TEVP_shear_stress_shear_rate}
\end{figure*}

\subsection{Stress relaxation regime}

In Fig.~\ref{fig::TEVP_param_shear_rate}, we compare the posterior distributions of the TEVP model's parameters obtained using observational data with $\dot\gamma_0 = 6~\si{\per\second}$, $7~\si{\per\second}$, and $8~\si{\per\second}$, to those obtained using observational data with $\dot\gamma_0 = 8~\si{\per\second}$, $9~\si{\per\second}$, and $10~\si{\per\second}$. Unlike the previous subsection, we now find that when data with $\dot\gamma_0 = 8~\si{\per\second}$, $9~\si{\per\second}$, and $10~\si{\per\second}$ is used for model parameters' calibration, the posterior distributions are narrower, and hence these shear rates are more informative for calibration of the TEVP model.

Finally, in Fig.~\ref{fig::TEVP_shear_stress_shear_rate}, we compare the propagated uncertainty of the shear stress profiles for two sets of observational data. Here, we again observe a much narrower HDI when using observational data for $\dot\gamma_0 = 8~\si{\per\second}$, $9~\si{\per\second}$, and $10~\si{\per\second}$, across the entire range of time. At lower shear rates, the noise in the experimental data evident in this figure, especially for $t<1~\si{\second}$ at $\dot{\gamma}_0=7~\si{\per\second}$, is appreciable, which naturally causes the posterior distributions to have a wider spread.

\begin{figure*}
    \centering
    \includegraphics[width=\linewidth]{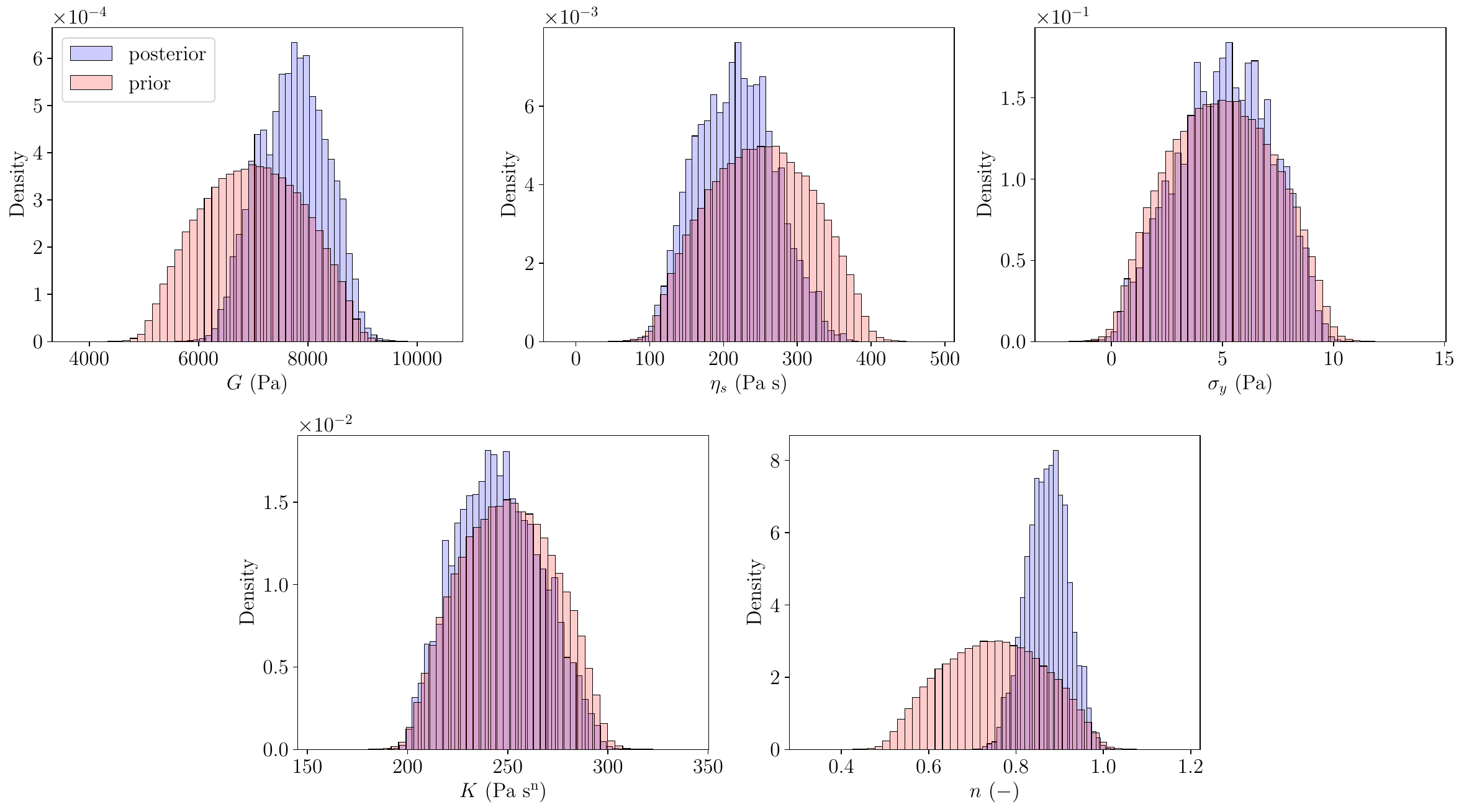}
    \caption{Prior and posterior distributions of the NEVP model's parameters.}
    \label{fig::prior_posterior_NEVP}
\end{figure*}

\begin{figure*}
    \centering
    \includegraphics[width=\linewidth]{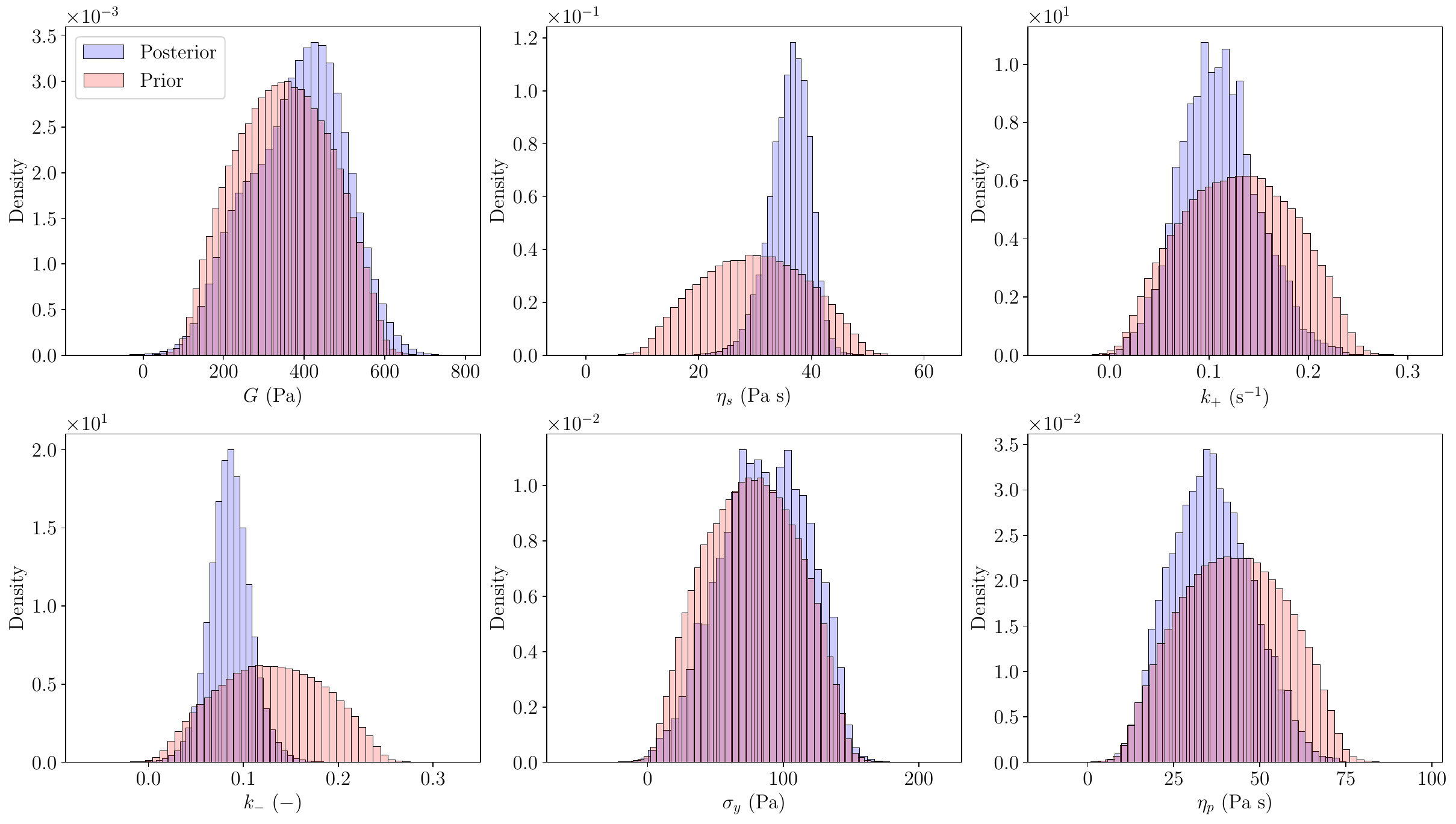}
    \caption{Prior and posterior distributions of the TEVP model's parameters.}
    \label{fig::prior_posterior_TEVP}
\end{figure*}

\section{Visualization of priors and posteriors of the models' parameters}
\label{app:prior_posterior}

In this appendix, we plot the priors and posteriors for the NEVP model's parameters in Fig.~\ref{fig::prior_posterior_NEVP} and TEVP model's parameters in Fig.~\ref{fig::prior_posterior_TEVP} on top of each other for easy visualization of the distributions. Typically, the priors have wider distribution (i.e., larger uncertainty) and are set by the analyst \emph{before} observing the data. Next, the developed hierarchical Bayesian inference approach outputs the posteriors of each model parameter from which we sample to obtain their distributions.

\bibliography{references.bib}

\end{document}